\documentclass[structabstract]{aa}  

\usepackage{graphicx}
\usepackage{txfonts}
\usepackage{natbib}
\bibpunct{(}{)}{;}{a}{}{,} 
\begin{document}
   \title{The Extended GMRT Radio Halo Survey I: New upper limits on radio
halos and mini-halos }

   \subtitle{}
\titlerunning{The Extended GMRT Radio Halo Survey I}
\authorrunning{Kale et al.}

   \author{R. Kale\inst{1,2}, T.Venturi\inst{2}, S. Giacintucci\inst{3,4}, D.
Dallacasa\inst{1,2}, R. Cassano\inst{2}, G.
Brunetti\inst{2}, G.
Macario\inst{5}, R. Athreya\inst{6}   
     }
\institute{Dipartimento di Fisica e Astronomia, Universita di Bologna, via
Ranzani 1, 40126 Bologna, Italy\\ \email{rkale@ira.inaf.it}
\and INAF-Istituto di Radioastronomia, via Gobetti 101, 40129 Bologna, Italy
\and Department of Astronomy, University of Maryland, College Park, MD
20742, USA
\and Joint Space-Science Institute, University of Maryland, College Park,
MD, 20742-2421, USA
\and Laboratoire Lagrange, UMR7293, Universit´e de Nice Sophia-Antipolis, CNRS,
Observatoire de la Cˆote dAzur, 06300 Nice, France
\and Indian Institute of Science Education and Research (IISER), Pune, India }

   \date{submitted to AA}

 
  \abstract
{ A fraction of galaxy clusters host diffuse radio sources called radio
 halos, radio relics and mini-halos. These are associated with the relativistic
 electrons and magnetic fields present over $\sim$ Mpc scales in the
 intra-cluster medium.}
  {We aim to carry out a systematic radio survey of all luminous galaxy clusters
selected from the REFLEX and eBCS X-ray catalogs with the Giant Metrewave Radio
Telescope, to understand the statistical properties of the diffuse radio
emission in galaxy clusters.}
   {We present the sample and first results from the Extended GMRT Radio Halo
Survey (EGRHS), which is an extension of the GMRT Radio Halo Survey
(GRHS, Venturi et al. 2007, 2008). }
  {Analysis of radio data at 610/ 235/ 325 MHz on 12 galaxy clusters are
presented. We report the detection of a newly discovered
 mini-halo in the cluster RXJ1532.9+3021 at 610 MHz. The presence of a small
scale relic ($\sim200$ kpc) is suspected in the cluster Z348. We do not detect
cluster-scale diffuse emission in 11 clusters.
Robust upper limits on the detection of radio
halo
of size of 1 Mpc are determined.   We also present upper
limits on the detections of
mini-halos in a sub-sample of cool-core clusters.
The upper limits for radio halos and mini-halos are plotted in the radio
power- X-ray luminosity plane and the correlations are discussed. Diffuse
extended
emission, not related to the target clusters, 
but detected as by-products in the sensitive images of two of the cluster 
fields (A689 and RXJ0439.0+0715) are reported.}
 {Based on the information about the presence of radio halos (or upper limits),
available on 48 clusters out of the total sample of
67 clusters (EGRHS+GRHS), we find that $23\pm7 \%$ of the clusters host 
radio
halos. The radio halo fraction rises to $31\pm11 \%$, when only the 
clusters
with X-ray luminosities $>8\times10^{44}$ erg s$^{-1}$ are considered. 
Mini-halos are found in $\sim50\%$ of cool-core clusters. A qualitative
 examination of the X-ray images of the clusters with no diffuse radio emission
indicates that a majority of
these clusters do not show extreme dynamical disturbances and supports the idea
that
mergers play an important role in the generation of radio halos/relics. The
analysis of the full sample will be presented in a future work.}


   \maketitle
%

\section{Introduction}
Galaxy clusters are massive ($\sim 10^{14} - 10^{15}$ M$_{\sun}$) 
assemblies of dark matter, diffuse gas and galaxies. The diffuse gas, called the 
intra-cluster medium (ICM),
mainly consists of hot  
thermal gas ($\sim 10^7 - 10^8$ K). It emits thermal Bremsstrahlung
radiation and is
detected in X-rays. Relativistic particles (with Lorentz factors $\gamma >
1000$) 
and magnetic fields ($\sim 0.1 - 1$ $\mu$G) are mixed with the 
thermal gas. The synchrotron emission associated with 
the ICM, detectable in the radio band, is a direct probe of the 
relativistic electrons and magnetic fields. 

The radio emission from the ICM has been classified into three main 
types: {\em radio halos, radio relics} and {\em mini-halos}
\citep[see][for a review]{fer12}.
Radio halos are $\sim$ Mpc sized sources having synchrotron spectra\footnote{$S
\propto \nu^{-\alpha}$, where $S$ is the flux density at frequency $\nu$ and
$\alpha$ is the spectral index.} with typical spectral indices, $\alpha \sim1.0
- 1.4$
and morphologies co-spatial with 
the extended X-ray emission from the ICM. 
Radio relics are diffuse radio sources 
at cluster peripheries, typically with filamentary, arc-like or irregular morphologies. 
Relics are also highly polarized and can have sizes in the range 0.2 to 2 
Mpc.
Mini-halos are 150 -- 500 kpc in size and have been found around 
 dominant radio galaxies at cluster centers.

Radio relics having elongated or arc-like
morphologies that are located at cluster peripheries,  
have been proposed to be produced by electrons that are accelerated by merger or
accretion shocks
\citep{ens98}. In some cases, the 
polarization properties and the spectral steepening from outer to inner edges
have been observed which confirms their connection  with the 
out-going
 cluster merger shock waves 
\citep[e.g.][]{gia08,wee10,kal12}. The Mach numbers of the shocks have
been estimated to be in the range $\sim 2 - 3$ based on the X-ray
observations \citep[see][for recent results]{aka11,aka11b,ogr12,sar12,bou13}.
Shocks
with such low Mach numbers are inefficient in accelerating
electrons to the observed energies and
the acceleration
mechanism behind relics is still not understood \citep[see][]{mar10,
jon12, kan12, pin13}. 

The origin of radio halos and mini-halos is still a matter of debate. The long
diffusion times of 
the relativistic electrons compared to their short radiative lifetimes require
an in-situ mechanism 
of generation of relativistic electrons in the ICM \citep{jaf77}. 

\noindent Mini-halos
have been found
in relaxed, cool core clusters \citep[e.g.][]{git02, gov09, gia11}.
Re-acceleration of a pre-existing population of
relativistic electrons in cool cores by turbulence has been proposed to explain
the mini-halos \citep{git02, maz08, zuh13}. 
The origin of turbulence in the ICM of cool
core clusters is still unclear, however it might be generated by gas sloshing
\citep{maz08}. The seed relativistic electrons may be injected 
from the activity of central AGN \citep[e.g.][]{cas08} and/or may be secondary
products of hadronic collisions in the ICM \citep[e.g.][]{pfr04, kes10}.

\noindent In the case of radio halos, there are two main classes of theoretical
models, namely, the `secondary electron' models and the turbulent
re-acceleration based models.
According to the secondary electron models, the relativistic electrons which
are produced as secondary products of the hadronic collisions in the ICM, in the
presence of magnetic fields lead to the generation of sources like radio halos
\citep[e.g.][]{den80,bla99,dol00,kes10}. 
The gamma rays expected from the 
hadronic collisions in the ICM have so far not been detected and this poses  
a challenge to the secondary electron models \citep[e.g.][]{jel11, ack10,
bru12}.

\noindent
In the turbulent re-acceleration based models, it is proposed that a low energy
relativistic electron
population in the ICM is
re-accelerated by turbulence
injected by mergers \citep{bru01,pet01,pet08}. The seed electron
population can be primary and/or secondary in origin.
The expectations of
these models, such as the occurrences
of radio halos in luminous, massive and merging clusters, have received support
from the observations \citep[e.g.][]{buo01,cas10}. According to these models the
most powerful radio halos are likely to occur in the energetic mergers
involving very massive galaxy clusters. A population of radio halos generated in
lower energy cluster mergers (more common), with a characteristic very steep
synchrotron
spectrum ($\alpha \geq 1.5$) due to lower energy in turbulence, is also
predicted \citep[e.g.][]{cas06}. These low energy
radio halos are termed as ultra-steep spectrum radio halos (USSRHs) and a
few have
been detected so far \citep[e.g.][]{bru08,mac10,mac13}. 
Due to their steep spectra,
instruments
operating at low frequencies such as the GMRT and the LOFAR are favoured for
detecting the USSRHs. The existence
of the USSRHs is also
a 
challenge for the secondary models that would require uncomparably large energy
content in cosmic ray protons \citep[see][and references
therein]{bru08}. Consequently LOFAR may allow a complementary test on the
origin of radio halos.

One of the most promising ways to make progress in the understanding of these
sources is to 
observe a large number of clusters to obtain statistics of the occurrence of 
radio halos, relics and mini-halos.
The success of this approach is evident from the results of the GMRT Radio Halo
Survey (hereafter GRHS)
 \citep[][hereafter, V07 and V08 respectively]{ven07,ven08}. The GRHS 
led to the discoveries of 4 radio halos, 1 mini-halo and 3 diffuse 
radio relic/halo candidates in galaxy clusters. 
Most importantly GRHS also gave the first upper limits on the detections of
radio
halos using the method of model radio halo injection.
The understanding of empirical 
correlations between the radio power at 1.4 GHz of the halo
(P$_{1.4\mathrm{GHz}}$) and 
X-ray luminosity (L$_X$) of the parent cluster 
and the connection between cluster mergers and radio halos 
have been significantly improved with the inclusion of upper limits. A limit on
the lifetime of a radio halo using the bimodal distribution 
of radio halo and non-radio-halo clusters in the P$_{1.4\mathrm{GHz}}$--L$_X$
plane 
 was obtained \citep{bru07}. The first quantitative estimate of the separation
of radio halo and 
non-radio halo clusters as merging and non-merging clusters has been made using
 the GRHS and the Chandra data \citep{cas10}. Recently, using the measurements
of integrated
Compton
 Y parameter of clusters with the Planck satellite, a scaling between radio halo
power and Y has been presented which shows a weaker bimodality as compared to
that in the P$_{1.4\mathrm{GHz}}$--L$_X$ plane \citep{bas12}. 
However, it is essential to
improve the statistical
significance of 
these relations to better understand the connection between the thermal and non-thermal 
components (relativistic electrons and magnetic fields) of the ICM. 

With this motivation we have undertaken the Extended GMRT Radio Halo Survey
(EGRHS). In this paper we present the first results from the EGRHS. The EGRHS
sample is described in Sec. 2.
The radio observations and data reduction are described in Sec. 3. The results
based on the radio
images are presented in Sec. 4. The estimates of the 
upper limits are presented in Sec. 5. The results are discussed in Sec. 6 and a
summary is presented in Sec. 7. Radio images of each of the cluster fields 
covering a region up to the virial radius are presented in Appendix 1.

A cosmology with $H_0 = 70$ km s$^{-1}$ Mpc$^{-1}$, $\Omega_m = 0.3$ 
and $\Omega_\Lambda = 0.7$ is adopted.

\section{The Extended GMRT Radio Halo Survey}
The EGRHS consists of a galaxy cluster sample
extracted from the 
ROSAT-ESO Flux Limited X-ray galaxy cluster catalog \citep[REFLEX,][]{boh04}
 and from the extended ROSAT Brightest Cluster Sample catalog
 \citep[eBCS,][]{ebe98,ebe00}. The selection criteria are :
\vspace*{-0.2 cm}
\begin{enumerate}
 \item L$_X$ (0.1-2.4 keV) $> 5\times10^{44}$ erg s$^{-1}$;
 \item $0.2 < z < 0.4 $;
 \item $\delta  > -30^{\circ}$ for the REFLEX and eBCS samples.
\end{enumerate}
 The declination limit, imposed while selecting
clusters from the eBCS catalog for the GRHS sample is extended to obtain the 
EGRHS sample.
The choice of high X-ray luminosity ensures that the radio halos (if present,
and with radio powers expected from the P$_{1.4\mathrm{GHz}}$-L$_X$ scaling)
will be well within the detection limits of the GMRT.
The X-ray luminosity and the redshift ranges also ensure higher possibility of 
the occurrence of radio halos based on model predictions \citep{cas04, cas05,
cas06}. 
The lower 
limit in declination of $-30^{\circ}$ will ensure good {\em uv-}coverage with
the 
GMRT. These selection criteria led to a sample of 67 clusters. 
Of these 50 clusters were part of the GRHS (V07, V08) and the additional 17
form the EGRHS (Table 1).

\section{Radio observations and data reduction}
The aim of the EGRHS and GRHS together is to assemble a large sample of galaxy
clusters with sensitive radio observations in order to improve the statistical
information on radio halos, relics and mini-halos. Therefore along with new
observations for EGRHS, we are also carrying out observations of a few GRHS
clusters for which radio data were inadequate (Table 1). Here we provide
a summary
of what is presented in this paper:
\begin{itemize}
 \item 7 clusters from EGRHS at 610 and 235 MHz (GMRT proposal codes 19\_039 and
16\_117);
\item 1 cluster from GRHS (A2261) at 610 and 235 MHz (GMRT proposal code
16\_117);
\item archival 325 MHz GMRT observations for 3 clusters in the GRHS (GMRT
Cluster
Key Science project, PI
V. K. Kulkarni) and
\item a re-analysis of 610 MHz data from V08 on the GRHS cluster RXJ1532.9+3021.
\end{itemize}
 The clusters and the corresponding frequencies of observation are
listed in
Table~\ref{t2}.

Based on the experience of observations at 610 MHz of GRHS clusters, we decided
to move to
dual frequency
(610-235 MHz simultaneous) observations for the EGRHS. This approach provides
 two major advantages as compared to the single frequency band  
observations of the GRHS. 
Firstly, it allows for an immediate cross check of the presence of diffuse
emission
at another frequency. Secondly, the low frequency (235 MHz) allows for a
possibility of the detection of USSRHs. In EGRHS each cluster is observed for
an 8 hour duration in the dual frequency mode. This has resulted in better {\em
uv-}coverage as compared to the GRHS. Thus the EGRHS is an extension of GRHS
with an upgraded observing strategy.

The dual frequency observations of the EGRHS were carried out with the GMRT
Software Backend \citep{roy10} in the mode that provided 256 channels to
acquire the data. 
Bandwidths of 32 MHz at 610 MHz and 8 MHz at 235 MHz 
were used. 

Data were analysed using the NRAO Astronomical Image Processing System (AIPS).
 The steps in data reduction that were followed at both 
610 and 235 MHz are described briefly.
The data from the antennas that were not working were identified by examining the 
calibrators and removed. The task `SPFLG' was used to identify and remove the 
channels affected by radio frequency interference. The channels at the edges of the 
bands were also removed due to lower sensitivity and stability. After excising
bad data, calibration using the 
primary and secondary calibrators was carried out. The calibrated data were 
re-examined and any 
remaining bad data were excised. The edited and calibrated data on target source 
were then averaged in 
frequency to an extent which kept `bandwidth smearing' effect at negligible levels.
The data were then imaged using wide field imaging technique. Several iterations of 
phase-only 
self-calibration and a final amplitude and phase self-calibration were carried out 
to improve the sensitivities of the images. Various tapers and weighting 
schemes on the uv-data were used to make images with a wide range of 
synthesized beams. Images with resolutions (FWHM) $4'' - 25''$ and $10'' - 
40''$ at 610 and 235
MHz, respectively, were examined. These were used in combination with high 
resolution images to test the presence of any suspected extended 
emission.

For two EGRHS clusters, Z348 and A267, and for the GRHS cluster A2261 
dual frequency data were
 recorded with the hardware backend. 
 It provided single polarization (RR) in two sidebands with 128 channels and a
bandwidth 
of 16 MHz each at 610 MHz. The 235 MHz data were recorded simultaneously in a
separate polarization (LL) in a single sideband with a bandwidth of 8 MHz.
The data were analysed separately for each sideband
and the images were then combined. In addition to these we also reanalysed the
610 MHz data on RXJ1532.9+3021 from V08.

Archival GMRT data at 325 MHz (bandwidth 32 MHz) were
analysed for three GRHS clusters, namely, 
RXCJ0437.1+0043, RXCJ1212.3-1816 and A2485 (GMRT cluster Key 
Project). The data analyses were carried out using the steps 
similar to the ones described above. 

\section{Results}
The analysis of these data have led to sensitive radio images of the
12 galaxy clusters. Radio images with rms noise in the range 45 - 80 $\mu$Jy
beam$^{-1}$ at 610 MHz, 0.25 - 0.4 mJy beam$^{-1}$ at 325 MHz and 0.45 - 1.8 mJy
beam$^{-1}$ at 235 MHz were obtained (Table 2). These are consistent with the
sensitivities obtained in the GRHS and thus ensure the
uniformity of the survey.

The main results of the paper are as follows. 
We present the 610 MHz image of the newly detected mini-halo in the cluster
RXJ1532.9+3021.
We did not detect radio halo, relic (Mpc scale) or mini-halo emission in 11
 clusters. Diffuse emission on smaller scales ($< 250$ kpc)
are suspected in Z348 and RXCJ0437.1+0043. We describe the images of individual
clusters in the following sub-sections.

\subsection{Radio images}
 The sensitivities achieved using 
`natural weights' (robust$= 5$ in AIPS) in each of the cluster fields are 
reported in Table 2. 
The residual
amplitude errors are $\sim5\%$ at 610 MHz \citep[e.g.][]{cha04} and $\sim5-15\%$
at 325 and 235 MHz. The radio images obtained using natural weights for the
{\it uv}-data (610/325 MHz) of central regions of the
clusters are shown in contours overlayed on the Chandra/XMM 
Newton\footnote{Data from Chandra observation IDs 7938, 10415, 9369, 3583, 
10440, 3580, 10465, 11729, 10439 and XMM Newton observation number 0652010201 
are used.} X-ray images shown in colour in Fig.~\ref{clus}. 
Radio images of all the
clusters covering regions up to the virial radius are presented in Appendix 1.
A short note on each of the clusters based on our radio images and the
information available in the literature is presented below.

\begin{figure*}
   \centering $
\begin{array}{ccc}
\includegraphics[height=4.5 cm]{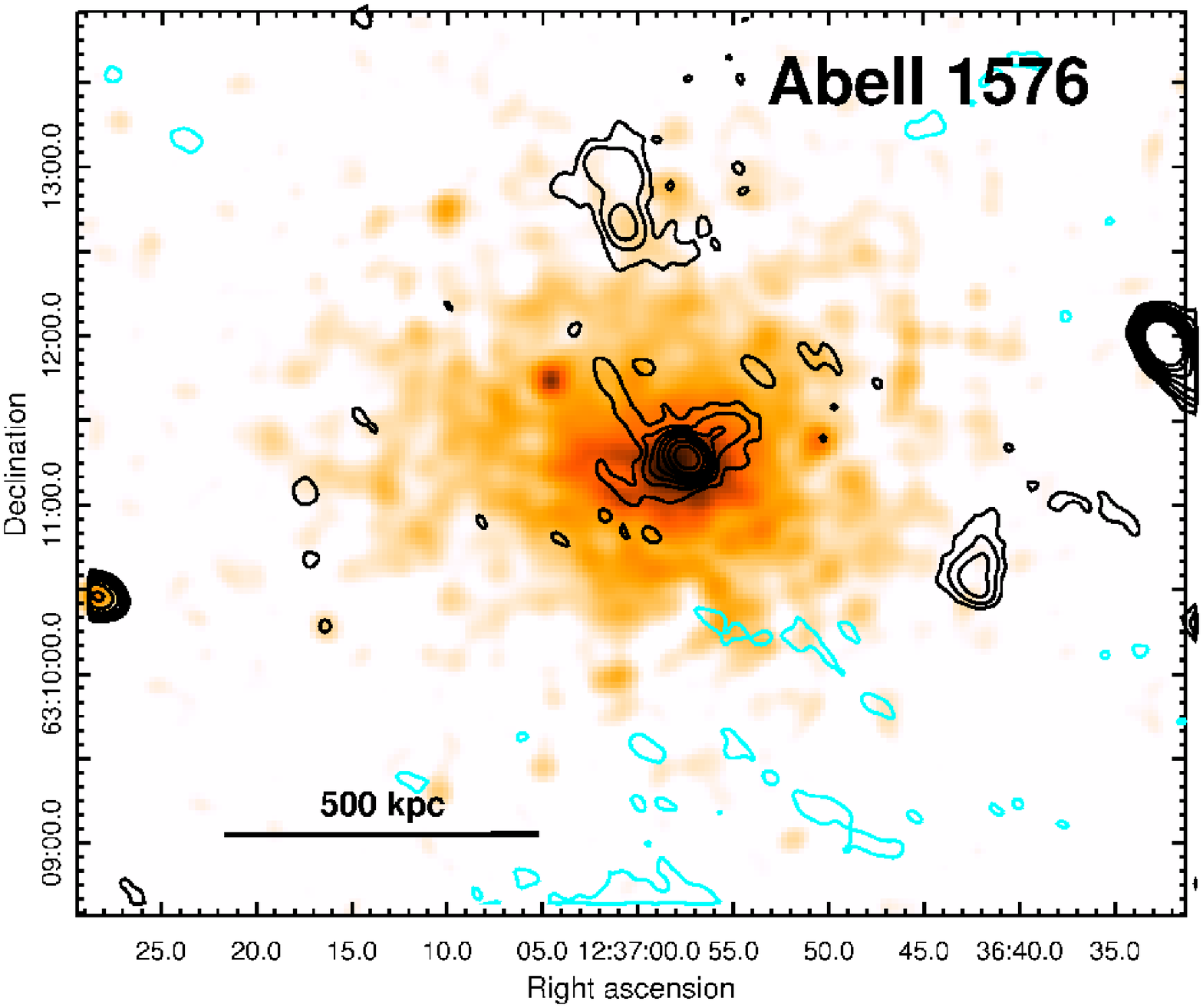}&
\includegraphics[height=4.5cm]{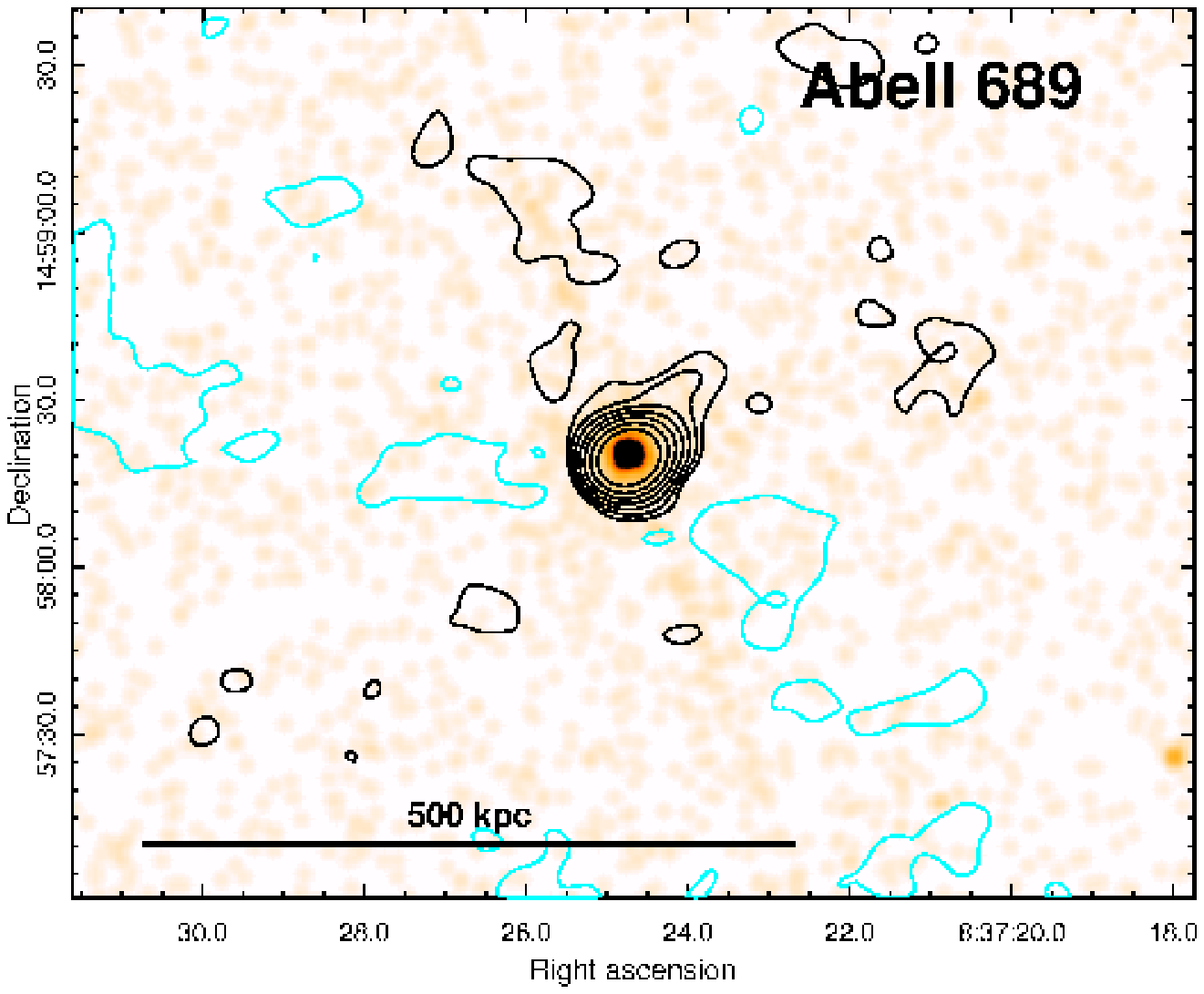}&
\includegraphics[height=4.5 cm]{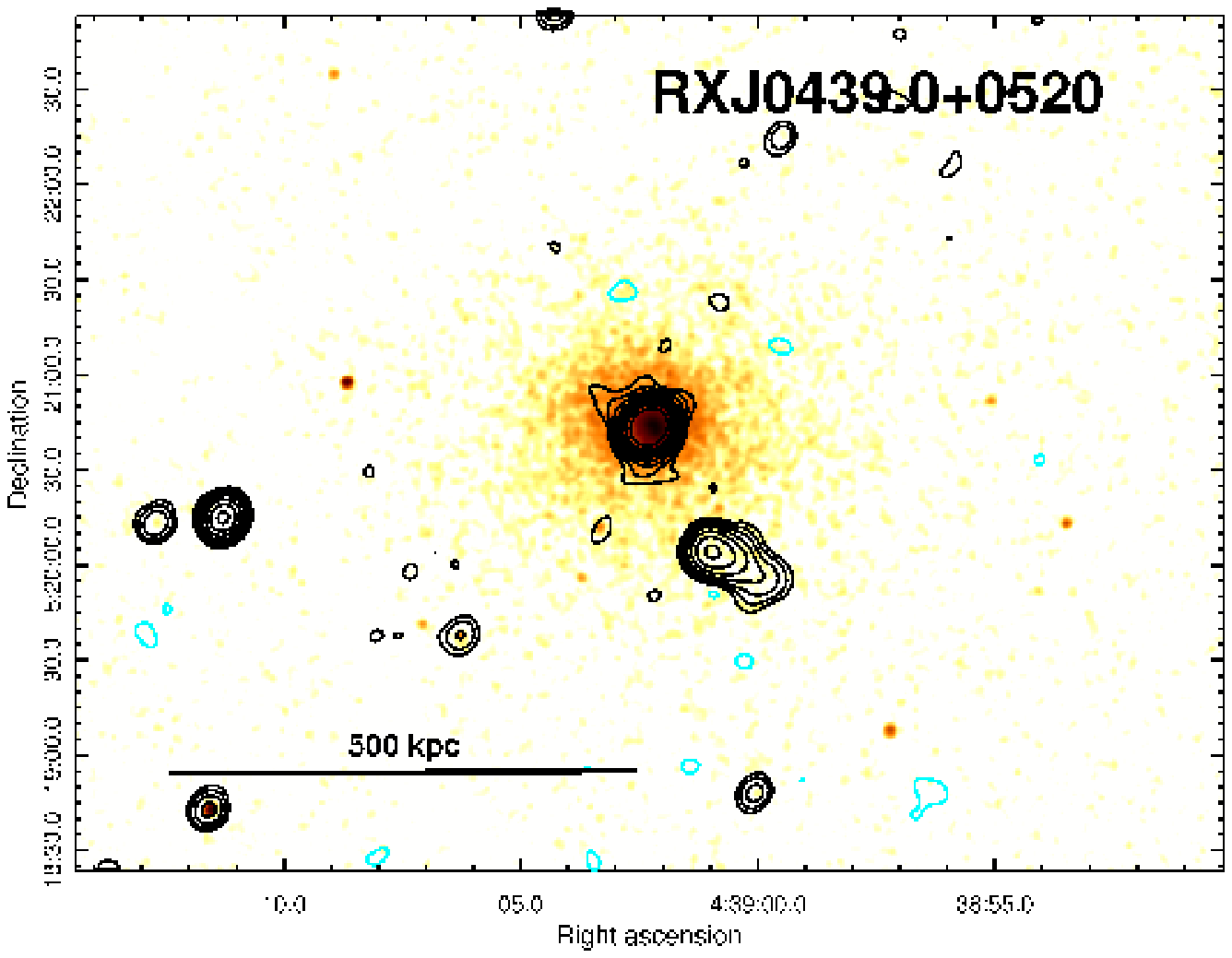}\\
\includegraphics[height=4.7 cm]{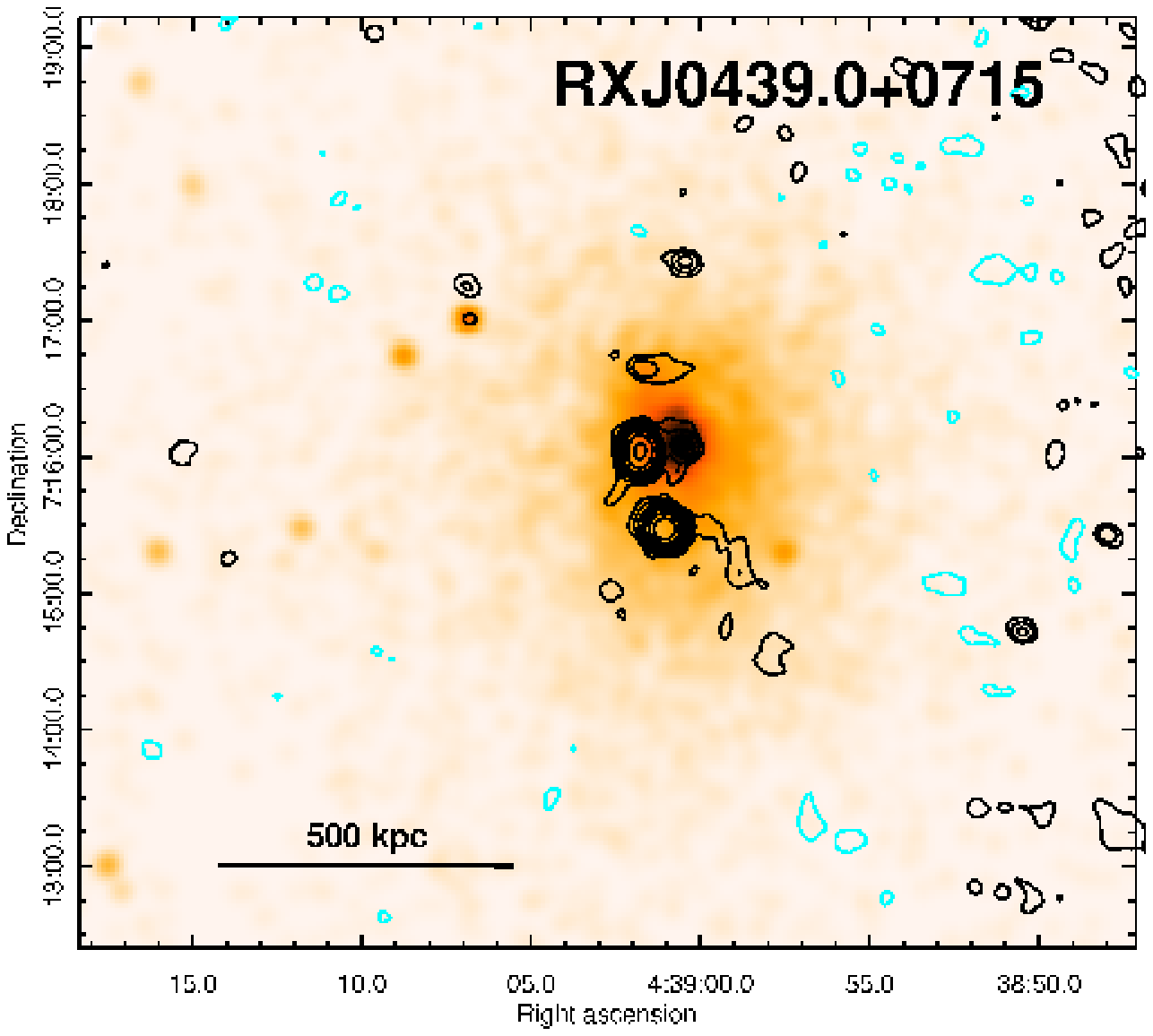} &
\includegraphics[height=5.1 cm]{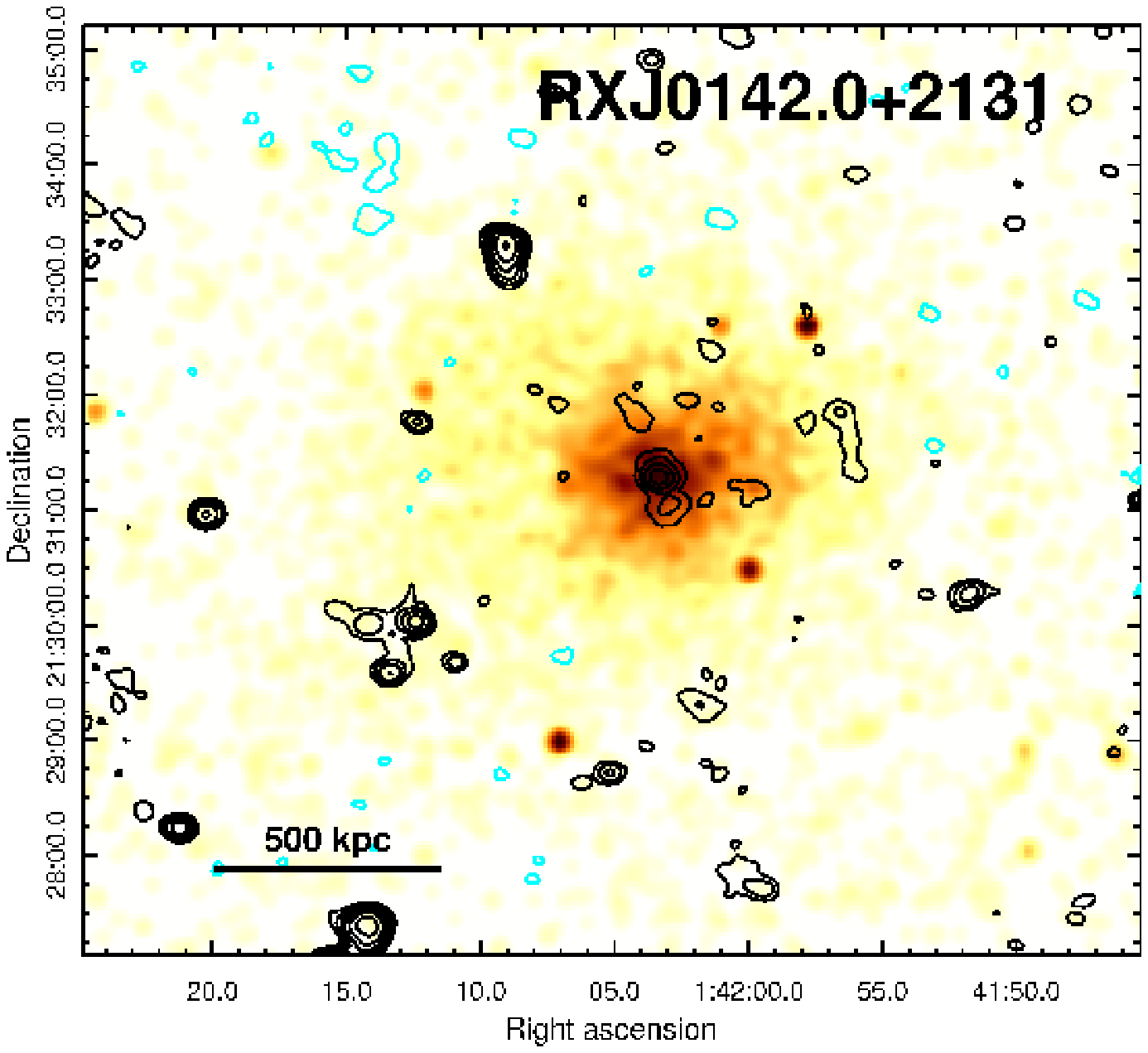} &
\includegraphics[height=4.6 cm]{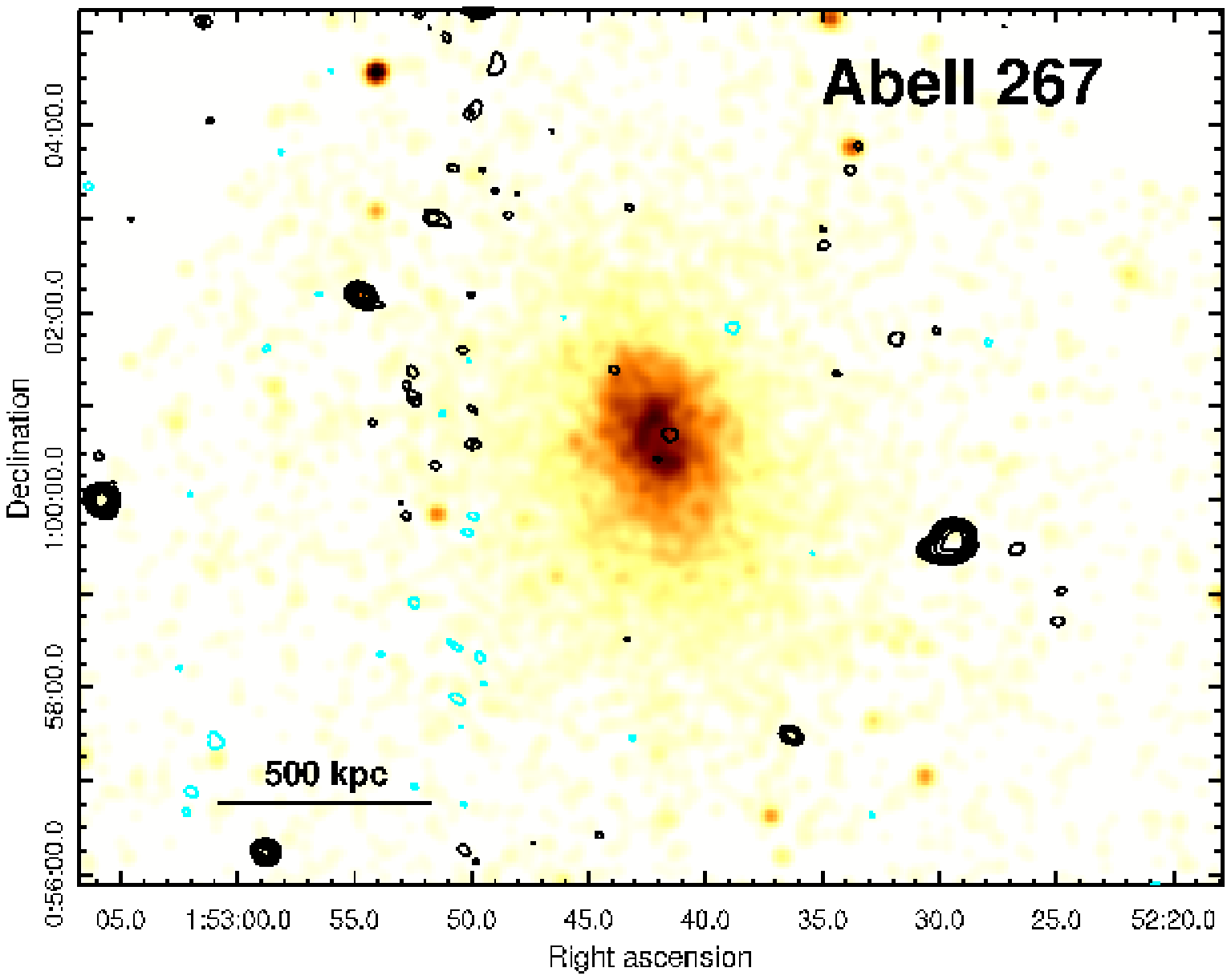} \\
\includegraphics[height=3.9 cm]{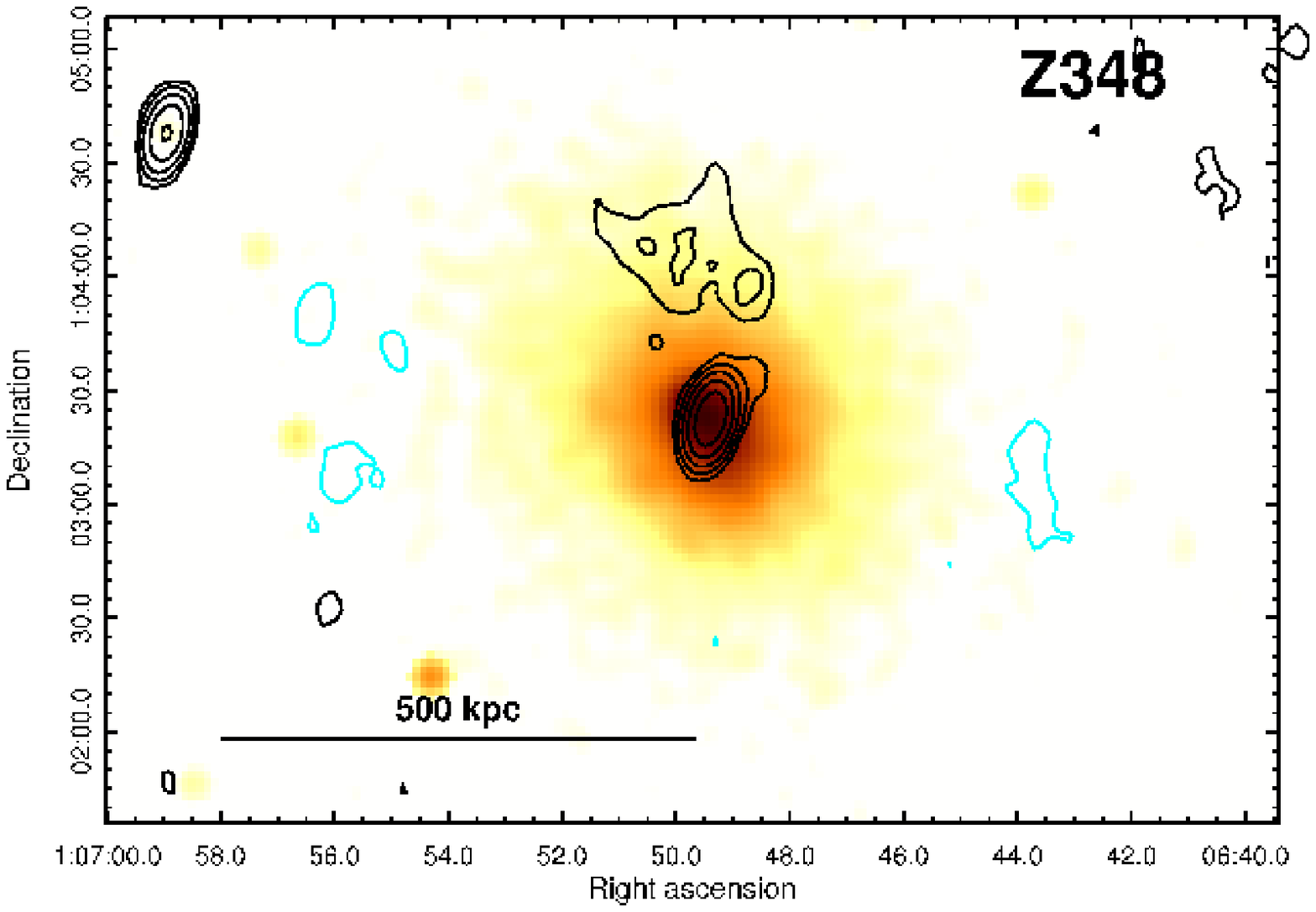} &
\includegraphics[height=4.7 cm]{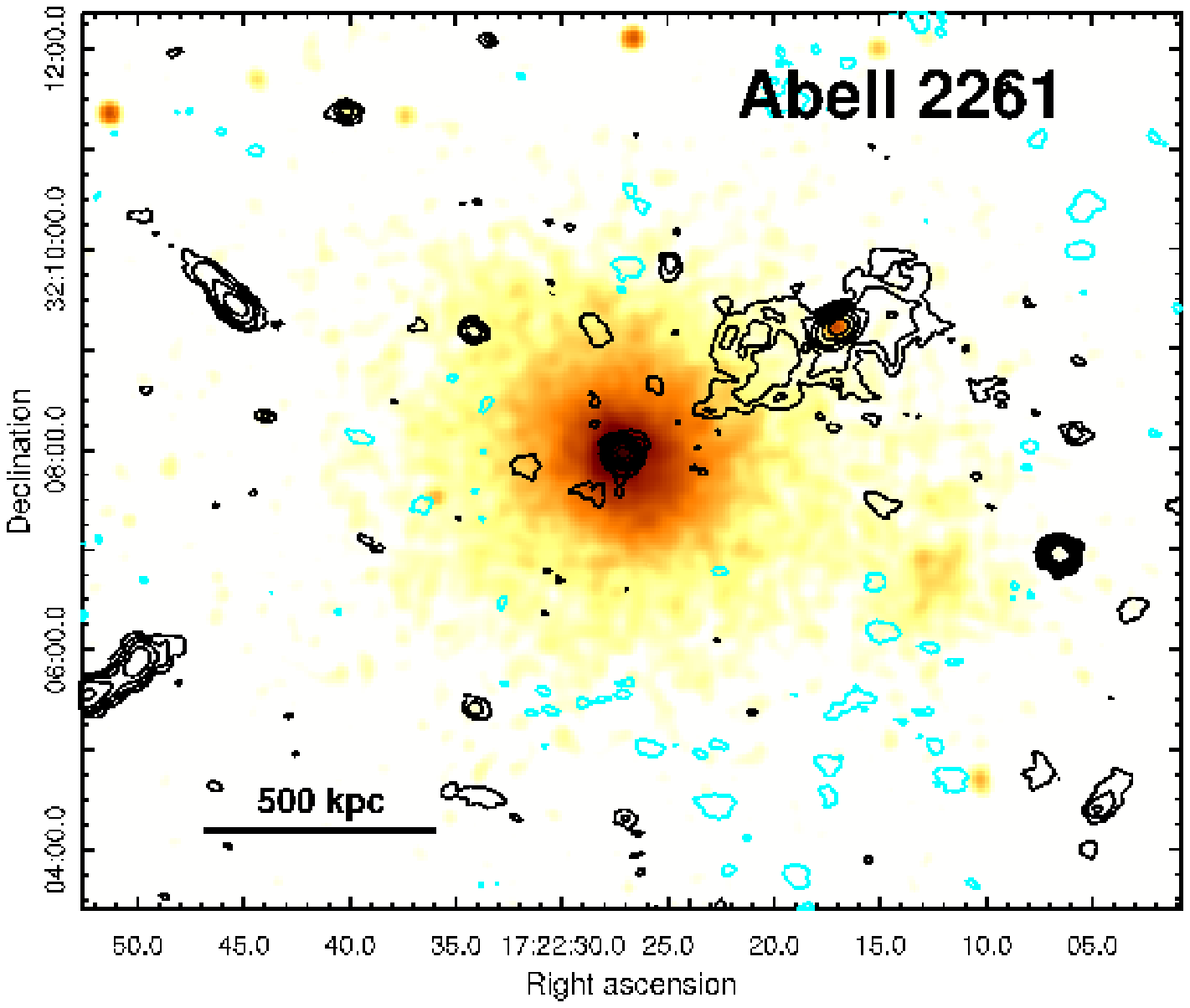}&
\includegraphics[height=5.0 cm]{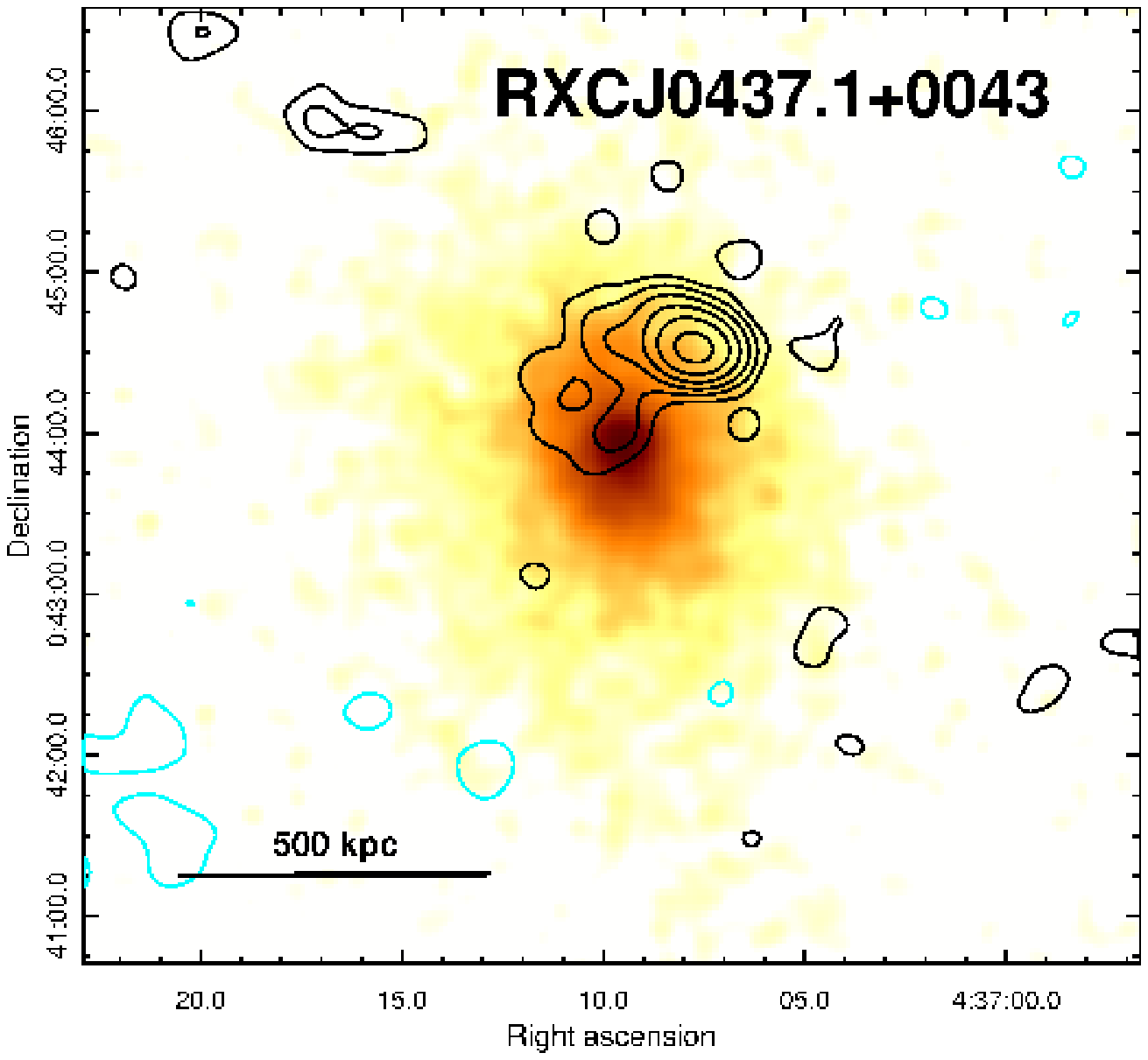} \\
\includegraphics[height=4.3 cm]{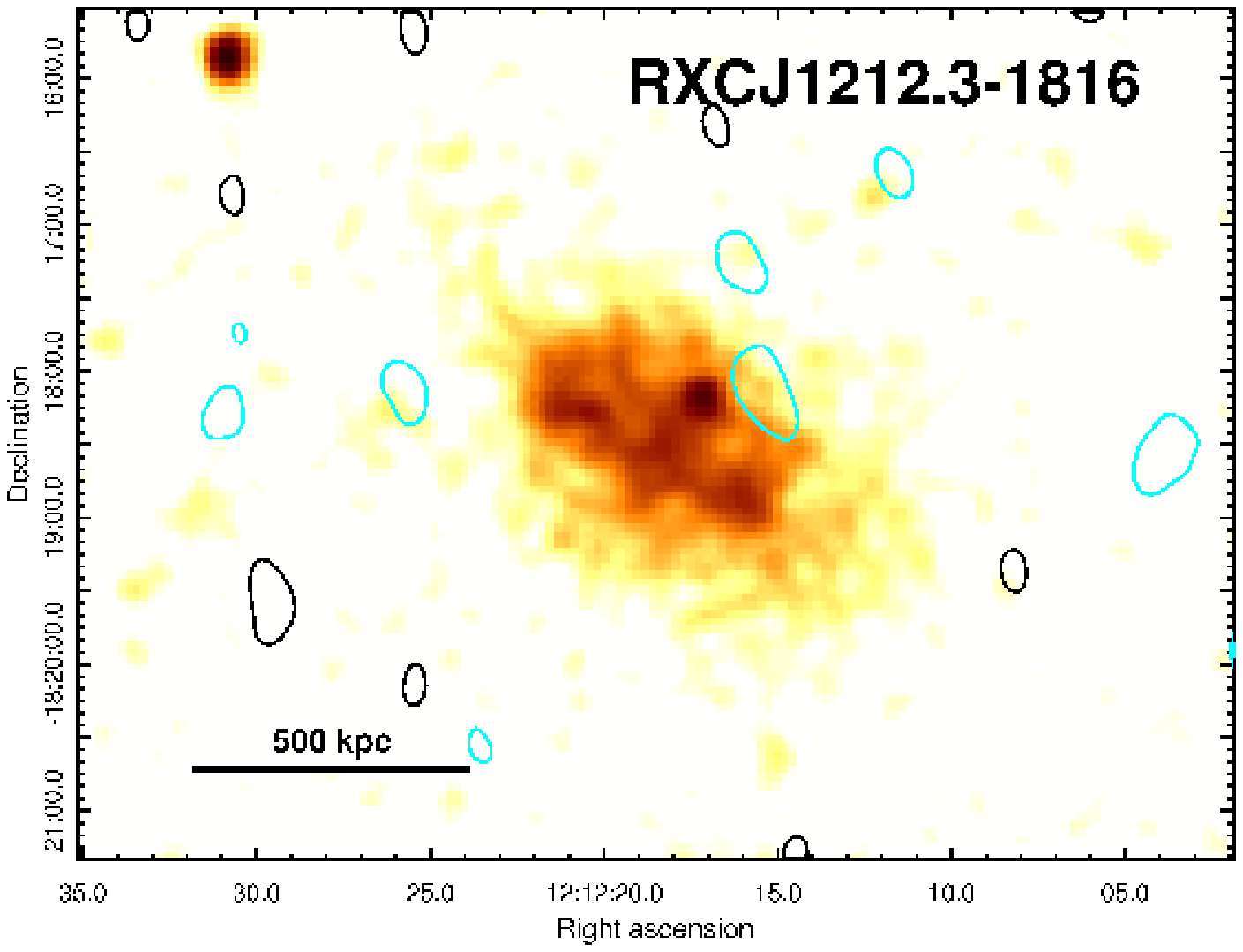} &
\includegraphics[height=4.5 cm]{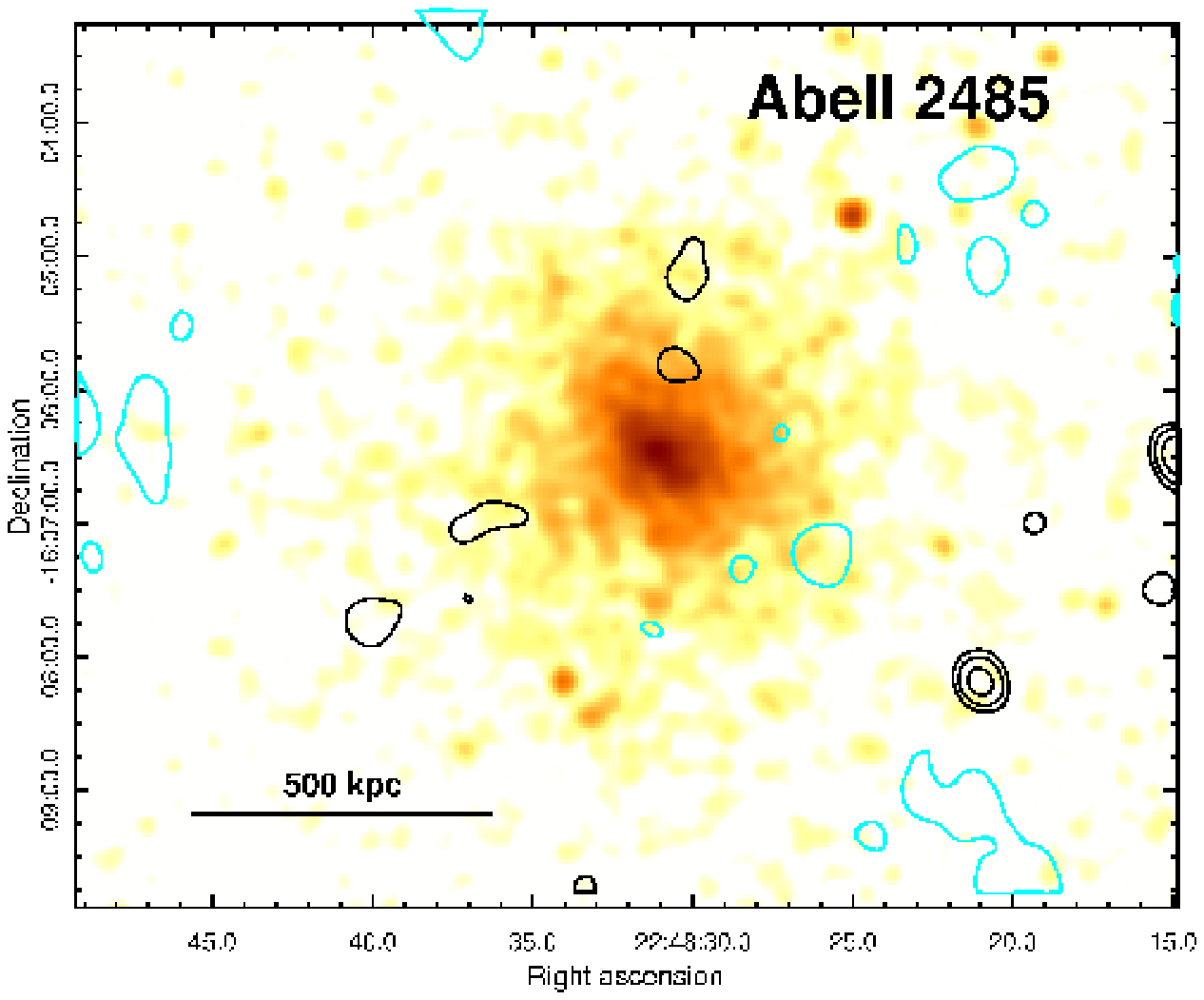}&
\end{array}
$
      \caption{Radio images (contours) are shown overlaid on X-ray images
(colour) for the 11 galaxy clusters presented in this paper. The radio images
are at 610 MHz for the clusters A1576, A689, RXJ0439.0+0520, RXJ0439.0+0715,
RXJ0142.0+2131, A267, Z348 and A2261 and at 325 MHz for the
clusters RXCJ0437.1+0043, RXCJ1212.3-1816 and A2485. The contour levels are at
$3\sigma\times(\pm1, 2, 4, 8, ...)$
(see Table 2 for $1\sigma$ levels and the beam sizes). Positive contours are
shown in black 
and negative in cyan. The exposure corrected Chandra (ACIS-I) images are 
presented for all clusters, except RXCJ1212.3-1816 for which XMM Newton (EPIC 
MOS) pipeline processed image is presented. The X-ray images are smoothed to 
resolutions of $10''-15''$.
}
         \label{clus}
\end{figure*}

\subsubsection{Abell 1576}
The cluster \object{A1576} (RXC J1236.9+631) is a rich cluster (richness, $R=3$)
at a
redshift of 0.302 ($1'' = 4.50$ kpc). The radio image shows the presence of a
central radio galaxy, almost coincident with the peak in the X-ray emission
(Fig.~\ref{clus}). In the highest resolution 610 MHz image (not shown), the
central source resolves into at least 2 components which are coincident with
optically detected galaxies in A1576. The optical counterpart of this radio
galaxy has at least
three optically detected nuclei and is suspected to be in an advanced stage of
merger \citep{dah02}. The central source has a jet-like extension 
toward the northwest. The slight extension toward northeast may be an artefact. 
The two other radio sources visible in the A1576 field shown in 
Fig.~\ref{clus} have optical counterparts and are possibly radio galaxies in
the cluster. 

The
field of this cluster contains several radio sources (Fig.~\ref{appa1576}). 
About $3'$ west of cluster center
is a double lobed radio galaxy. The core of this galaxy
is not detected 
in the 610 and 235 MHz radio images and also in the FIRST survey. Thus the
optical counterpart for this radio galaxy is
not obvious. The morphology of the lobes indicates 
that it is of FR-II type. The spectral indices (235 - 610 MHz) of the
eastern and the western lobes are 0.65 and 0.72, respectively. 

Based on weak
lensing analysis, \citet{dah02} infer significant dynamical activity in A1576.
The X-ray surface brightness distribution is elongated in the east-west
direction
and does not show pronounced central peak (Fig.~\ref{clus}). The cluster is
relatively hot, with a
temperature of 8.65 keV \citep{cav09}.

\subsubsection{Abell 689}
\object{A689} (RX J0837.4+1458) is a cluster at a redshift of 0.279 ($1''=
4.26$ kpc)
with richness,
$R=0$. There is a bright X-ray and radio source associated with the central 
AGN in A689 (Fig.~\ref{clus}). 
The positive and
negative emission at $\sim 3\sigma$ level, surrounding the bright
central source are noise structures. 
The radio image of A689 contains several
bright sources (Fig.~\ref{appa689}). In particular there is a complex blend of
radio
galaxies and a low surface brightness feature toward the southeast of the
cluster center (Fig.~\ref{appa689}, see Sec.
4.2.2). 

Prominent sub-structures have been reported in the mass distribution 
derived from
lensing measurements \citep{oka10}. 

\subsubsection{RXJ0439.0+0520}
\object{RXJ0439.0+0520} is a luminous cluster at redshift 0.208 ($1''=3.41$
kpc). 
There is a radio source at the center and  another source
toward its south, possibly a head-tail galaxy (Fig.~\ref{clus}). The central
source has some hint
of extended emission (on $\sim 50$ kpc scale), but needs higher resolution
observations to confirm it.
About $4'$ east of the center, a possible FR-I type radio galaxy is seen
(Fig.~\ref{app5rx}). A bright radio 
galaxy with a distinct core and two extended lobes with hot spots is detected
about $9.6'$ west of the center (Fig.~\ref{app5rx}). Optical counterparts to 
both these radio galaxies are detected in DSS POSS-II images. However, their
connection to the galaxy cluster is unclear due to unavailability of redshifts
for
the optical counterparts.

The cluster has a temperature of 4.63 keV \citep{cav09} and a circularly
symmetric morphology in X-rays \citep{jel05} (Fig.~\ref{clus}). This implies
that this cluster is most likely relaxed, with no recent dynamical
activity.

\subsubsection{RXJ0439.0+0715}
The cluster \object{RXJ0439.0+0715} is located at a redshift of 0.244
($1''=3.86$ kpc). 
A weak radio source is coincident with the peak in the X-rays
(Fig.~\ref{clus}). Adjacent to this source is a bright radio source that
resolves into a double source, oriented in the north-south direction, in
the high resolution image at 610 MHz (not shown).
In the cluster field up to the virial radius, a bright radio galaxy
 $\sim7'$ north-west of the cluster center is seen (Fig.~\ref{app7rx}). A
pattern due to
improper deconvolution of this bright source is visible
in Fig.~\ref{app7rx}. The tail-like extension of the radio source, just south
of the cluster center is most likely a part of this pattern.

The cluster is elliptical and elongated along north-south direction in X-rays
(Fig.~\ref{clus}). The cluster has a temperature of 6.5 keV \citep{cav09}.
 There is no information about the dynamical state of this cluster in the
literature.

\subsubsection{RXJ0142.0+2131}
\object{RXJ0142.0+2131} is a cluster at a redshift of 0.280 ($1''=4.27$ kpc). 
 A radio source is present at the
cluster center, coincident with the peak of the X-ray emission
(Fig.~\ref{clus}). The cluster field has several bright
radio sources
(Fig.~\ref{apprx0142}). 

An amorphous distribution of X-ray emission is seen in the Chandra image with
no strong central peak (Fig.~\ref{clus}). This cluster
has a velocity dispersion of  $1278\pm134$ km s$^{-1}$, however, no
substructure in the velocity distribution \citep{bar05}. Based on the
mass-to-light ratios of galaxies and $\alpha-$element abundance ratios, a
possibility of a past merger in the cluster has been inferred \citep{bar06}. 

\subsubsection{Abell 267}
The cluster \object{A267} is at a redshift of 0.230 ($1''=3.69$ kpc) with
richness,
$R=0$.
A faint radio source is detected at the center of this cluster
(Fig.~\ref{clus}). The cluster field has some compact bright radio sources
(Fig.~\ref{appa267}). 

The cluster has a cD galaxy at the center \citep{dah02} and the X-ray
distribution is elliptical \citep{jel05} (Fig.~\ref{clus}). The cluster is
moderately hot with a temperature of
8.7 keV \citep{cav09}. It has been classified as a cool core cluster based on
the analysis of power ratios \citep{bau05}. However, its temperature profile
lacks the typical drop in the centers of cool-cores and it has a relatively
high value of central entropy ($170$ keV cm$^{2}$; \citet{cav09}). This
central entropy is much higher than 50 keV cm$^{2}$, which is the value adopted
to separate cool cores and non-cool cores \citep[][]{cav09, ros11}. Therefore
A267 is like a non-cool core based on its entropy and temperature, but a
relaxed cluster based on the power ratios.

\subsubsection{Z348}
\object{Z348} (RXC J0106.8+0103) is at a redshift 0.254 ($1''=3.98$ kpc). A
compact
radio source is detected at the center of this cluster (Fig.~\ref{clus}). 
Adjacent to the central source, $\sim 30''$ toward north, a diffuse source is
detected in the 610 MHz image (Fig.~\ref{clus}). The source is elongated in the
east-west direction and has an extent of $\sim 200$ kpc, if assumed to be at the
redshift of Z348. It could be a relic of smaller extent similar to the
one in A85 \citep{sle01}. The nature of this source remains to be confirmed. 

There is an AGN at the cluster center with strong X-ray emission
\citep{boh00}, which is also detected as a compact radio source
(Fig.~\ref{clus}). 
The Chandra X-ray image shows elliptical distribution of emission along the
northeast-southwest
direction. No information about the dynamical status of Z348 is available in the
literature.

\subsubsection{Abell 2261}
\begin{figure}
   \centering
   \includegraphics[height = 6 cm]{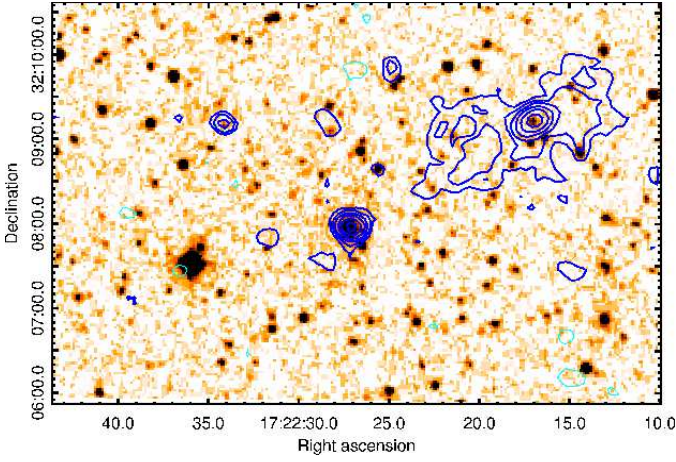}
\caption{GMRT 610 MHz image of the central region of the cluster A2261 shown in
contours overlayed on DSS POSS-II R band optical image shown in colour. The
contour levels and the beam are the same as in Fig.~\ref{clus}. An
optical source is coincident with the central radio source and with
the core of the radio galaxy.}\label{a2261opt}
\end{figure}

\object{A2261} is a rich cluster ($R=2$) at a redshift of 0.224 ($1''=3.61$
kpc). There
is a central radio source in the cluster (Fig.~\ref{clus}) identified with the
optical galaxy 2MASX J17222717+3207571. A remarkable radio galaxy is seen
($\sim2.5' = 540$ kpc) toward the northwest of the cluster center. The radio
galaxy shows presence of a compact source at the center and diffuse emission
extended on two sides of it (Fig.~\ref{a2261opt}). The diffuse
emission is filamentary and of total extent $\sim590$ kpc (if at the redshift
of the cluster). Based on the morphology it is possibly an `FR-I'
type radio galaxy. The total flux density
(core$+$diffuse) of the radio galaxy at 610 MHz is $57\pm6$ mJy and 23.3 mJy at
1.4
GHz (NVSS J172216+320910). The implied spectral index is 1.08. The compact X-ray
source, \object{CXOGBA J172217.0+220913} \citep{gil09}, and an uncatalogued
galaxy in the SDSS r-band image is co-incident with the compact radio core
(Fig.~\ref{a2261opt}). The field of this
cluster shows the presence of several radio galaxies (Fig.~\ref{appa2261}).

Diffuse radio emission was suspected in this cluster based on the 1.4 GHz VLA D
array
image (V08). However, we did not find cluster wide extended radio emission at
610 and 235 MHz.

The cluster has circular morphology in X-rays and is most likely a relaxed
 system. It has a temperature of 7.58 keV \citep{cav09}.
There is a diffuse patch of X-ray emission, west of the cluster,
however its connection to the cluster is unknown.

\subsubsection{RXCJ0437.1+0043}
The cluster \object{RXCJ0437.1+0043} is at a redshift of 0.284 ($1''=4.31$ kpc).
There
is a weak radio source at the cluster center and an elongated radio source
to the north
of it (Fig.~\ref{clus}). The extent of the elongated source in the east-west
direction is $\sim 256$ kpc, if assumed to be at the redshift of the cluster. 
These sources were also detected at 1.4 GHz by \citet{fer05}. Their flux
densities at 325 MHz are $4.0\pm0.5$ mJy (central source) and $55\pm5$ mJy
(elongated source). The spectral indices for the two sources between 325 MHz
and 1.4 GHz are 0.8 and 1.0, respectively. The elongated source is suspected to
have an extension detected at 325 MHz toward the east which is
not detected at 1.4 GHz. Both of these sources have possible optical
counterparts but need to be confirmed.

A surface brightness upper limit of 0.18 mJy beam$^{-1}$ (beam $=
23''\times21''$, p. a. $23^\circ$) at 1.4 GHz based on the $3\sigma$ level in
the image was reported by \citet{fer05}. An upper
limit using the method of injection of radio halo is presented in this work
(Sec. 5).

This cluster
field has other bright radio sources (Fig.~\ref{apprx0437}). A double radio
galaxy is located $\sim5.2'$ east of the cluster center.

The X-ray emission from the cluster appears elliptical and elongated, roughly in
the north-south direction (Fig.~\ref{clus}). The cluster has a temperature of
7-8
keV in the
outer regions and 5 keV at the center and is most likely a relaxed
cluster \citep{fer05}.

\subsubsection{RXCJ1212.3-1816}
\object{RXCJ1212.3-1816} is at a redshift of 0.269 ($1''=4.14$ kpc). There is no
radio
source in the vicinity of the cluster center (Fig.~\ref{clus}). The cluster
field shows a few bright radio sources (Fig.~\ref{apprx1212}).

The cluster
appears
elongated in the northeast-southwest direction in the X-ray image with XMM
Newton (Fig.~\ref{clus}). The morphology also hints at a disturbed ICM.  There
is no information in the
literature about the dynamical status of this cluster.

\subsubsection{Abell 2485}
\object{A2485} is at a redshift of 0.247 ($1''=3.89$ kpc). The 610 MHz data on
this
cluster was presented in V08. However, it was not considered for analysis due
to high rms noise (V08). Radio image at 325 MHz is presented here. There is no
bright radio
source close to the cluster center (Fig.~\ref{clus}). The field up to virial
radius shows presence of several compact radio sources (Fig.~\ref{appa2485}).

The X-ray emission may be
disturbed as the central peak is not pronounced (Fig.~\ref{clus}).
There is no information about the dynamical status of this cluster in the
literature.

\subsection{A mini-halo in RXJ1532.9+3021}
\begin{figure}
   \centering
   \includegraphics[height = 8 cm]{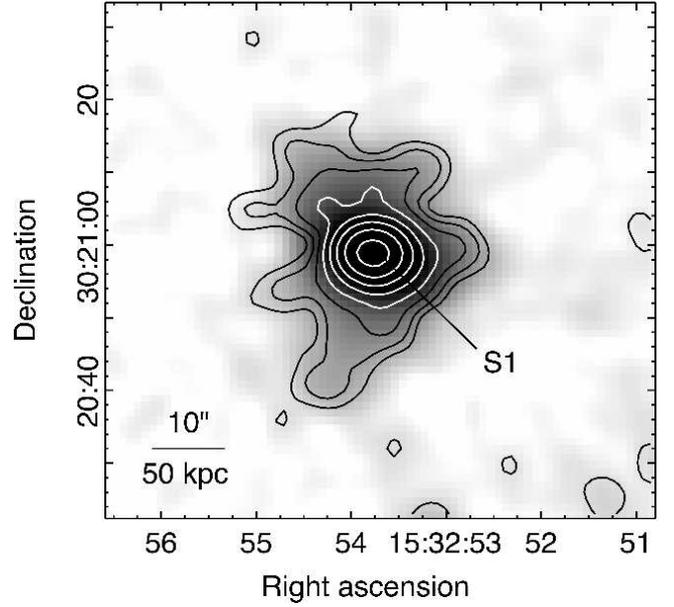}
\caption{GMRT 610 MHz image of RXJ1532.9+3021 shown in contours
and grey-scale. The contours levels start at 0.12 mJy beam$^{-1}$ and
increase by a factor 2. The beam is
$5.6^{\prime\prime}\times5.1^{\prime\prime}$, p.a.
$82^{\circ}$. The rms noise and the beam in the grey-scale image are $50\mu$Jy
beam$^{-1}$ and $9.2^{\prime\prime}\times6.6^{\prime\prime}$, p.a.
$69^{\circ}$ respectively. A compact source S1 is at the center
and is surrounded by the mini-halo.}\label{rx1532}
\end{figure}

The cluster \object{RXJ1532.9+3021} is at a redshift of 0.345 ($1''=4.86$ kpc). 
We present the 610 MHz image of a mini-halo discovered in this cluster using
VLA 1.4 GHz data (Giacintucci et al. in prep.). The 610 MHz observations of this
cluster were presented in V08, however the
mini-halo was not noticed. We re-analysed the 610 MHz observations from V08 and
obtained radio 
images, shown in Fig.~\ref{rx1532}, that clearly reveal the diffuse mini-halo
around the central unresolved radio galaxy (marked S1). The flux density of the
mini-halo is $\sim16$ mJy (after subtraction of the compact source S1, whose
flux density is $\sim 20$ mJy) at 610 MHz. The largest linear size of $\sim 200$
kpc is detected at 610 MHz. The properties of this mini-halo will be
reported in detail elsewhere (Giacintucci et al. in prep.). 

Based on the X-ray properties, this cluster is classified as a cool core
cluster \citep{lar11}. It has an average temperature of 5.65 keV \citep{cav09}.
\subsection{Extended emission associated with galaxies}
The EGRHS and the GRHS surveys are designed to be sensitive to low surface
brightness, extended features such as the radio halos and relics. The resulting
sensitive 
images thus also have the potential to detect other kind of faint extended
emission 
such as synchrotron emission from galaxies, faint portions of radio lobes or 
even relic lobes of radio galaxies. We present the detection of such emission
associated with galaxies found in the images of the clusters A689 and
RXJ0439.0+0715.
\subsubsection{A disk galaxy and a radio galaxy with faint lobes}
\begin{figure*}
   \centering
   \includegraphics[height=7 cm]{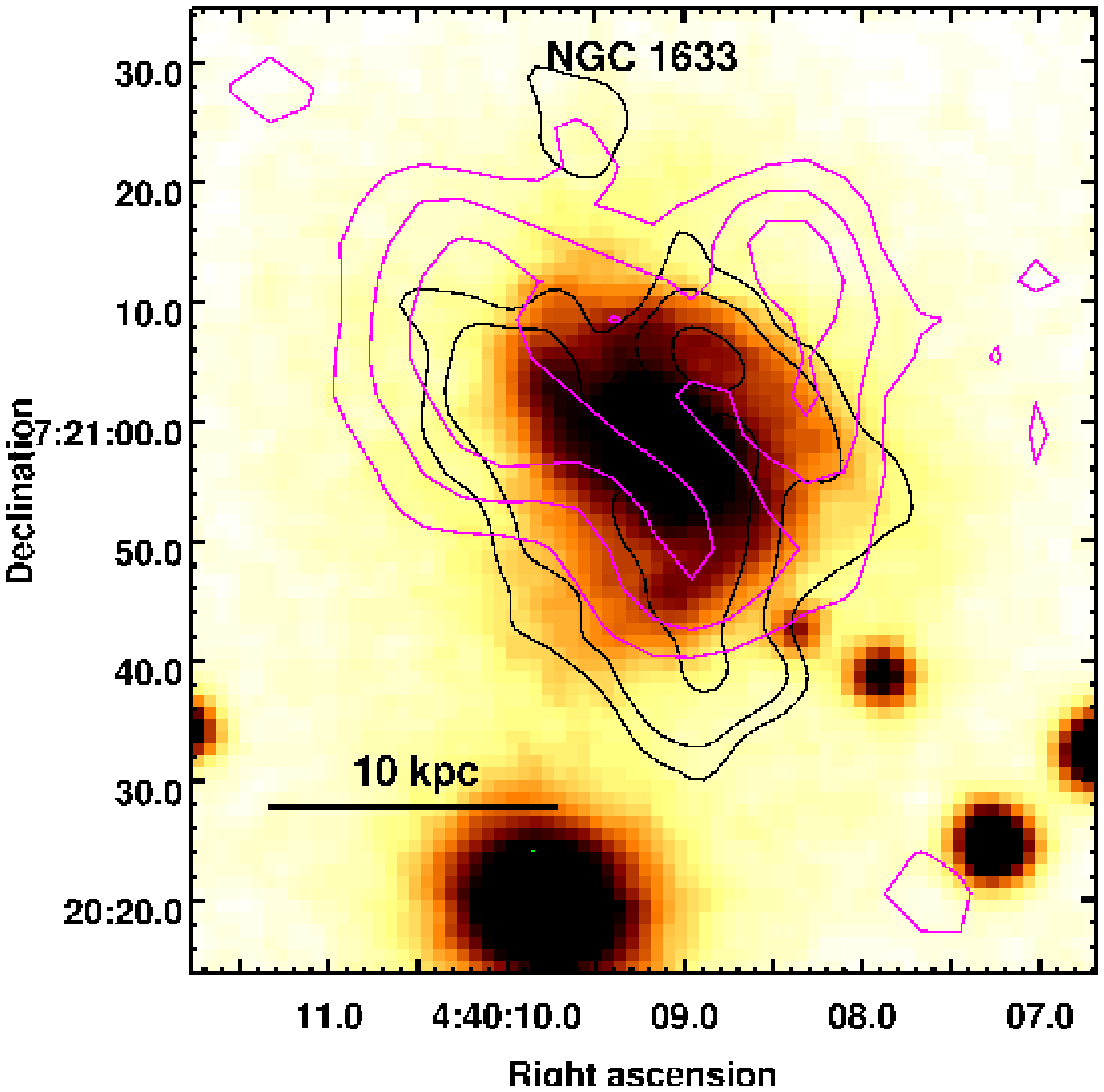}
 \includegraphics[height=7 cm]{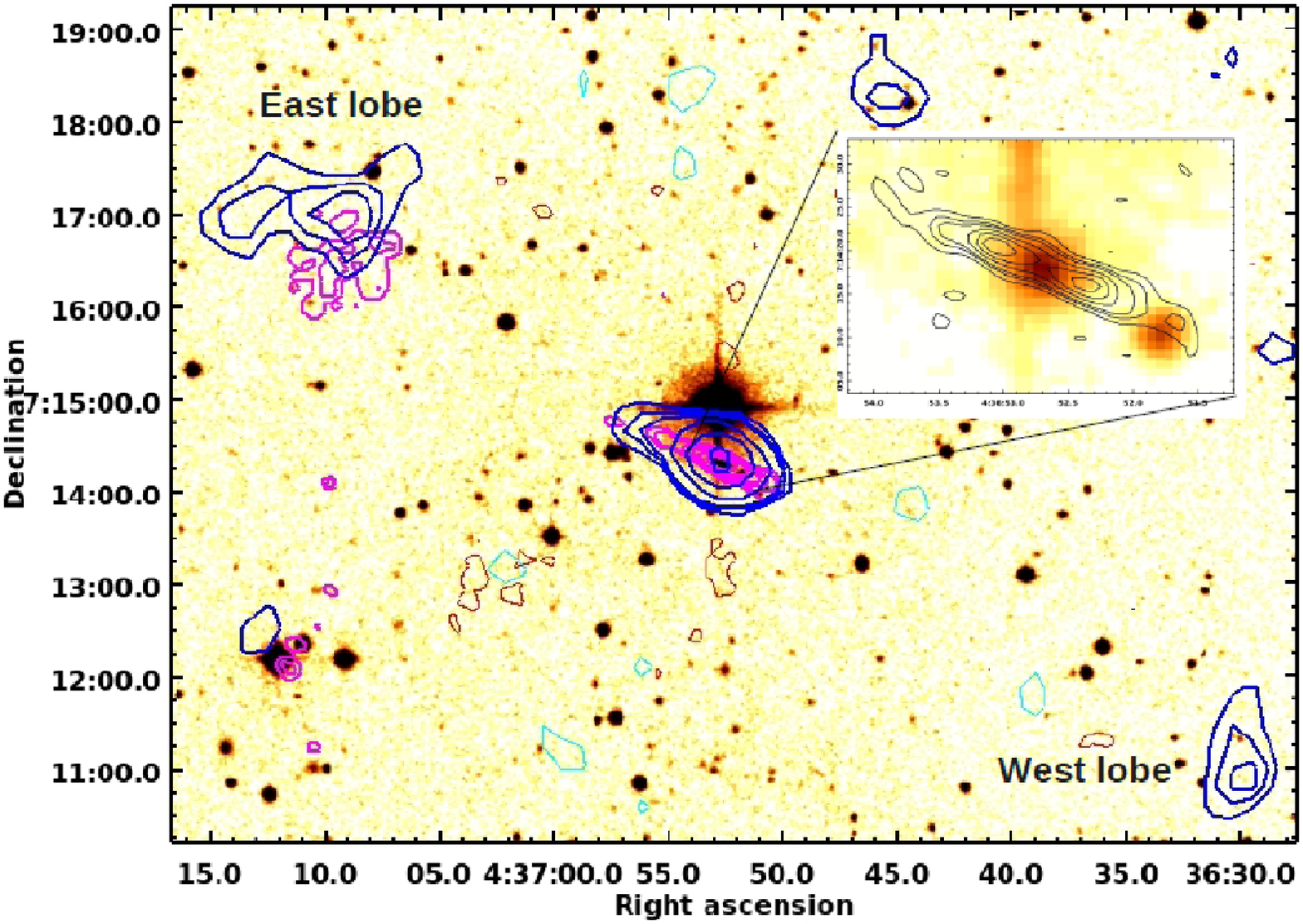}
      \caption{{\it Left} Radio emission from the spiral galaxy NGC 1633 
(DSS R band in colour) detected at 610 MHz 
(black contours) and 235 MHz (magenta contours). Contour levels at 
610 MHz are -0.2, 0.2, 0.3, 0.45 mJy beam$^{-1}$.
{\it Right} A possible `double-double' radio galaxy with the outer lobes
(labeled
east 
and west lobe) and the inner lobes shown in the inset. The 610 MHz contours
(positive 
magenta and red negative contours) and 235 MHz contours (positive blue 
and cyan negative contours) are overlayed on UKIRT infra-red image shown in 
colour. Contour levels at 610 MHz are -0.18, 0.18, 0.24, 0.36, 0.48, 0.72, 1.44 
mJy beam$^{-1}$ and at 235 MHz are -1.5, 1.5, 2.0, 2.5, 3.0, 6.0, 12.0 
mJy beam$^{-1}$.
The black contours in the inset are of the high resolution 610 MHz image 
that resolves 
the core into lobes with contour levels at -0.15, 0.15, 0.3, 
0.4, 0.6, 0.8, 1.0 mJy beam$^{-1}$.}
         \label{7rx}
\end{figure*}

In the field of view of the cluster RXJ0439.0+0715, $\sim17'$  
east of the center, a patch of diffuse emission was detected at 
610 and 235 MHz. The NVSS survey also detects this source, however, with a
resolution
of $45''$, it is not resolved.
Overlay of the radio image on the Digitized Sky Survey R-band image revealed the
presence 
of the galaxy \object{NGC1633} at location of the diffuse patch 
(Fig.~\ref{7rx}, left). \object{NGC1633} is a spiral galaxy  at a redshift
of 
0.016642 (De Vaucouleurs et al. 1991, RC3.9). Since there is no compact 
source detected in higher resolution images made at 610 and 235 MHz, 
this diffuse emission is most likely synchrotron emission from the disk of the
galaxy. The 
largest linear size of the emission detected at 610 MHz is $\sim$16 kpc. 
The flux densities at 610 and 235 MHz estimated using primary beam corrected 
images are $13.4\pm2.0$ mJy and $29.0\pm4.0$ mJy, respectively. A spectral 
spectral index of 0.8 is implied.

About $32'$ west of the center of this cluster field another interesting 
source was noticed. The NVSS source NVSSJ043652+071420, is 
resolved at 610 and 235 MHz to reveal a radio galaxy with jets oriented 
in the direction northeast -- southwest (Fig.~\ref{7rx}, right).  
The core is coincident with a galaxy detected in the UKIRT Infra Red Survey
 (Fig.~\ref{7rx}, right). The flux densities at 610 and 235 MHz of the core
region are $78\pm7$ mJy and $69\pm5$ mJy, respectively, indicating a
self-absorption.  
Also along the same northeast -- southwest direction, two diffuse sources 
are detected in the 235 MHz image. We propose that these are `lobes' associated
with the
radio galaxy. The angular separations of the `lobes' from the core are 
$4.8'$ (eastern lobe) and $6.6'$ (western lobe).
The eastern lobe is also detected in the 610 MHz image. 
The western lobe is at the edge of the 610 MHz field of 
view, in a region with poor sensitivity, and thus cannot be detected. 
No jets connecting these lobes to the core are detected.
These lobes are not detected in the NVSS due to their low brightness. 
The high resolution
(uniform weighted) images at 610 and 235 MHz resolve out the lobes, 
confirming their diffuse nature. There are no obvious optical or infra-red
counterparts to these.
The diffuse lobes, due to their locations relative to 
the central double, are likely to be lobes of the radio galaxy, 
also possibly from a previous activity.
We did not find any identification or redshift information of the 
 host galaxy. 

\subsubsection{A complex blend of radio sources and a galaxy filament?}
Toward the southeast of the center of the field A689, two 
radio sources with complex morphologies were detected. 
A complex blend of radio sources and a filament of diffuse emission are detected
$\sim9'$ and $\sim15'$ southeast of the cluster center, respectively
(Figs.~\ref{appa689}). 

\begin{figure*}
   \centering
  \includegraphics[height=6.5 cm]{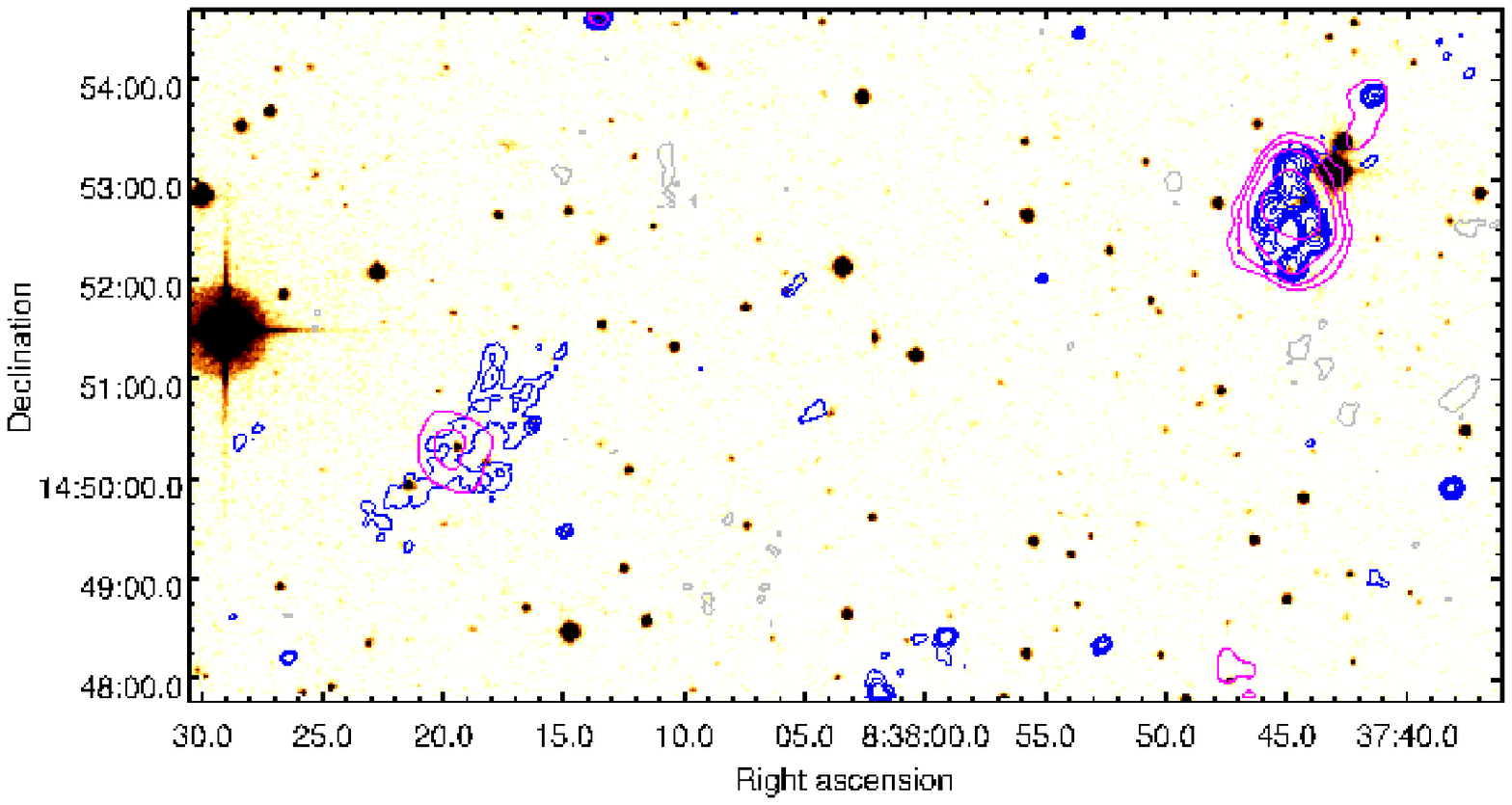}
 \includegraphics[height=6.5 cm]{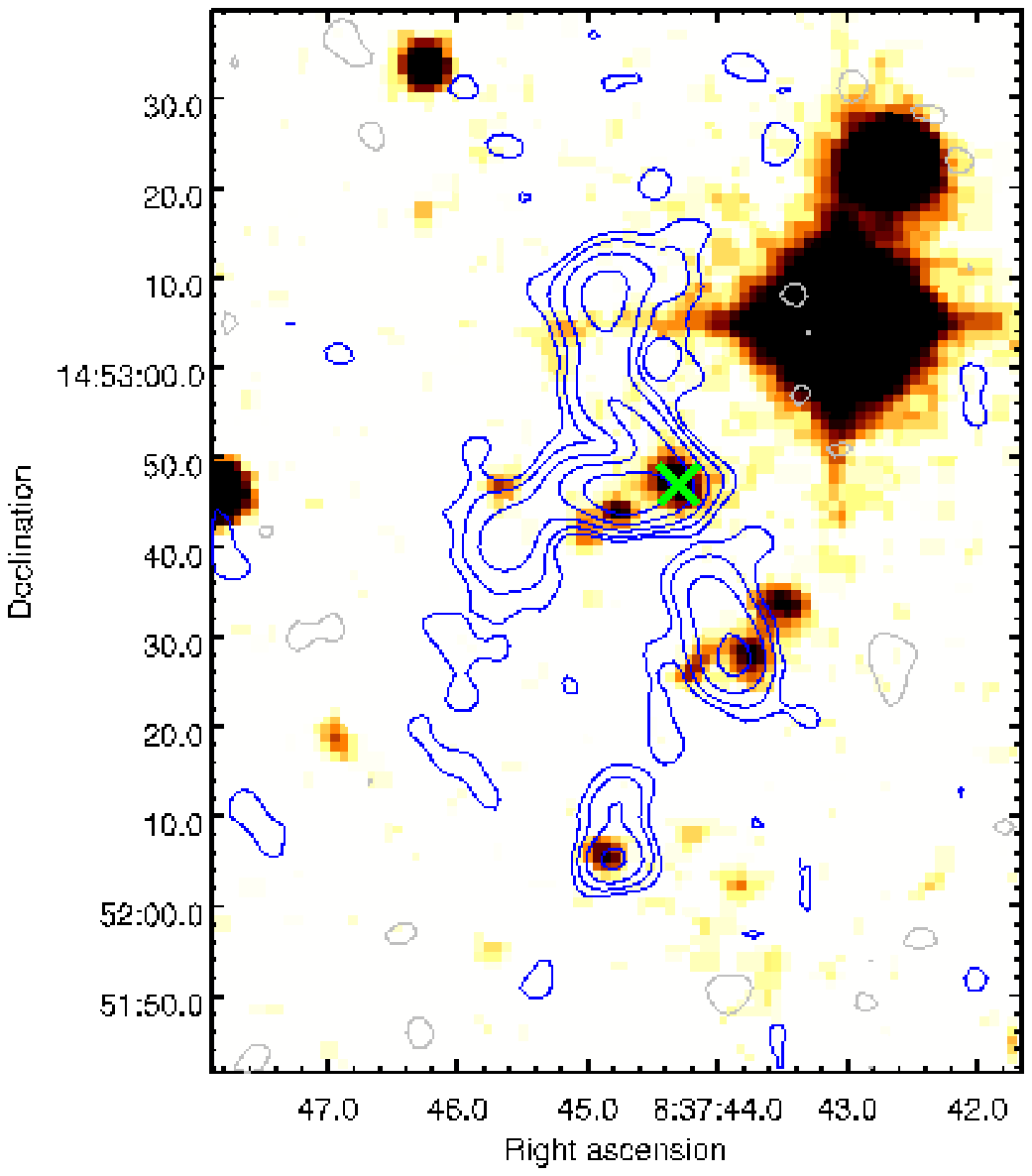}
      \caption{{\it Left} The complex blend of radio galaxies is toward the
right and
the diffuse emission is toward left shown in contours at 610 and 235 MHz on the
DSS R band
image in colour. Contours at 610 MHz are $0.2\times(\pm1, 1.5, 2, 4, 8, 16,...)$
mJy beam$^{-1}$. The positive contours are in blue and negative in grey.
The magenta contour levels at 235 MHz are $3.0, 4.0, 6.0, 12.0$ mJy beam$^{-1}$.
There are no
negative contours at 235 MHz in this region. The synthesized beams are
$7.7''\times6.2'', \mathrm{PA}~ -63.6^{\circ}$ at 610 MHz and
$29.7''\times25.2'', \mathrm{PA}~ 35.7^{\circ}$ at 235 MHz. {\it Right} The
complex blend of radio galaxies in the high resolution 610 MHz image. The
synthesized beam is $4.3''\times3.2''$, PA -80.0. The contour levels are
$0.15\times (\pm1,2,4,8,...$) mJy beam$^{-1}$. The position of the galaxy
cluster WHL J083744.3+145247 is marked by a `$\times$'.}
         \label{a689}
   \end{figure*}

The complex blend of radio sources consists of three distinct radio galaxies
(Fig.~\ref{a689}). A galaxy cluster 
 \object{WHLJ083744.3+145247} with a photometric redshift 0.2875 
 is co-spatial with these galaxies \citep{wen09, wen10}. No X-ray source is
known 
at this location. The hosts of the radio galaxies are likely members of this
cluster.
The northernmost source is a narrow angle tailed (NAT) radio galaxy. There
are two other radio sources with unresolved (cores) and resolved components
(tails) to the south of the NAT (Fig.~\ref{a689}, right). A bridge of radio
emission connecting the
southern source 
and the eastern tail of the NAT is also seen at both 610 and 
235 MHz. The bends in the jets of these radio galaxies in the north or
northeast direction may be indicators of the ram pressure exerted by the
surrounding medium. 

The diffuse emission further southeast of the 
radio galaxy complex is detected in both 610 and 235 MHz images
(Figs.~\ref{a689}, ~\ref{appa689}). 
The source \object{NVSS J083816+14505} is co-spatial with the northwest portion 
of the diffuse emission. A few galaxies
are detected in the Sloan Digitized Sky Survey (SDSS) at the location of the 
diffuse emission. It is possible that the diffuse emission is a blend of the
emission from these individual galaxies. No redshift information is available 
for the SDSS galaxies. Deeper radio observations of this region are required to
map the morphology of this emission. Determination of the redshifts of the
galaxies in this region will enable to estimate the linear extent of the
diffuse emission.

\section{Upper limits}
Firm upper limits on the detections of radio halos in these clusters are
 an important part of this survey. With the non-detections of radio halos in
the 11 clusters, we
proceeded to the step of determining the upper limits. 
We followed a procedure for placing firm upper
limits on the flux density of extended 
emission that was used in the GRHS \citep[][V08]{bru07}. The procedure consists
of 
introducing simulated 
(fake) radio halos of a given size and brightness in the {\em uv-}data and 
then re-imaging the data. Use of radio images over a wide range of resolutions  
 is made in establishing the upper limit. The typical rms noise in low 
resolution images is reported in Table 3.

\subsection{Radio halos}
A fake radio halo of 1 Mpc diameter, which is the
typical size of giant radio halos, was chosen for injection. It  
was modeled using optically thin concentric spheres to match the average profile 
of well studied radio halos \citep[][V08]{bru07}. 
The task `UVMOD' was used to add the model to the {\em uv-}data. The new 
{\em uv-}data were used to make image and the detection of the fake radio 
halo was examined in it. Fake radio halos with flux densities at 610 MHz 
over a range of 3 - 20 mJy were injected in the data sets. For those with 
data at 325 MHz the flux densities of the injected halos were scaled with 
a spectral index of $1.3$. The upper limits 
obtained on the detection of a 1 Mpc diameter radio halo 
are listed in Table 3. Following the earlier works (V08), a spectral index of
$1.3$ was assumed to estimate the radio upper limit at 1.4 GHz.

\begin{figure}
   \centering
\includegraphics[height=8.8 cm]{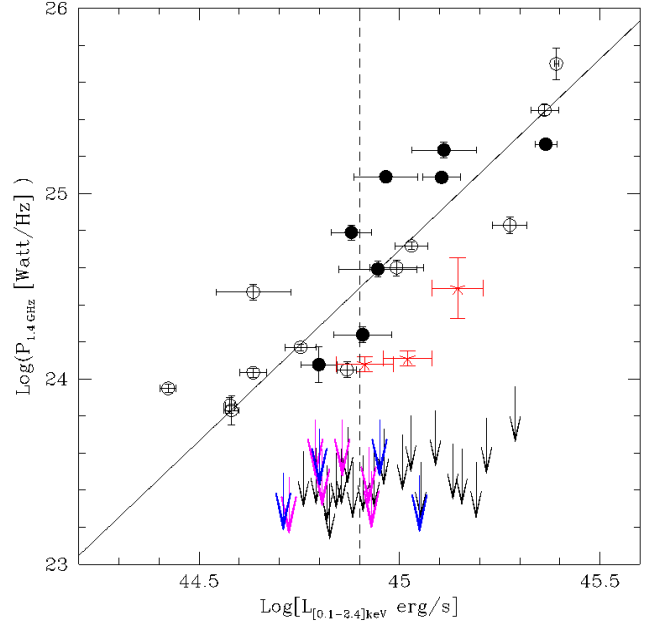}
      \caption{ The new upper limits (magenta and 
blue arrows) are shown in the $P_{1.4\mathrm{GHz}}$ - $L_X$ plane with the 
correlation line (black solid line) for giant radio halos reproduced from 
\citet{bru09}. The magenta 
arrows are EGRHS clusters, the blue arrows are
GRHS clusters presented here (see Table 1). The giant radio halos that are 
hosted in 
the clusters in the EGRHS+GRHS sample are denoted by filled black circles and 
red asterisks (USSRHs). The other giant radio halos (black open circles) and 
 upper limits (black arrows) are reproduced from \citet{bru09}.}
         \label{lxlr}
\end{figure}

 The upper limits obtained here are plotted  
in the P$_{1.4\mathrm{GHz}}$ -- $L_X$ plane 
(blue and magenta arrows, Fig.~\ref{lxlr}). The old upper limits (black 
arrows), the giant radio halos (filled and open black points) and the 
correlation 
line obtained from these giant radio halos shown in the plot are reproduced 
from \citet{bru09}. 
The red asterisks are the 
known ultra-steep spectrum giant radio halos in the GRHS sample, namely, A697 
\citep[V08][]{mac10}, A521 \citep{bru08, dal09} and A1300 \citep{ven13}. The 
filled points, the red points and the upper limits together provide a view of 
the GRHS+EGRHS sub-sample in this plane.

 From the known radio halos, there are indications that the more powerful 
radio 
halos tend to also be the most extended in linear size 
\citep{cas07,mur09,gio09}. Of course the morphologies of radio halos are 
complex in most cases and the measure of the extent of radio halo is 
not robust. We obtained radio halo size as expected from 
the relation between the halo radius and the expected radio power in order to 
carry out injections. These expected sizes differed from the size of 1 
Mpc by $2- 22 \%$ with only 3 clusters showing differences $>10\%$. In the 
three extreme cases, namely, RXJ0439.0+0520, A2261 and A2485, we injected radio 
halos with the expected sizes of 0.84, 1.22 and 0.82 Mpc respectively. The 
upper limits in A2261 and A2485 are 8 mJy and 5 mJy respectively. The upper 
limit in RXJ0439.0+0520 did not change.
We report the upper limits obtained using 1 Mpc injections in Fig. 
~\ref{lxlr} and Table 3. 

\subsection{Mini-halos}
Radio mini-halos are diffuse radio sources found in a fraction of cool-core
clusters, surrounding the central galaxy. We extracted cool core clusters from
the GRHS and EGRHS samples. 
The cool core clusters were
identified using conditions based on central entropy ($K_0 < 50$ keV cm$^2$),
cooling time ($t_{cool} < 2$ Gyr) and luminosity ratio 
($L_{core}/L_{500}$\footnote{Ratio of core X-ray luminosity to the
luminosity within $R_{500}$ (Cassano et al. in prep.).} $> 0.5$). The central
entropy and
cooling time limits adopted here are standard criteria to identify cool cores
\citep[e.g.][]{bau05, cav09, ros11}. The luminosity ratio is an 
additional property which is often used to identify cool core clusters 
\citep[e.g.][]{san08, cas10}. Five clusters from the GRHS, namely,
RXJ1532.9+3201, A2667,
RXCJ1115.8+0129, Z2089 and Z2701, and two clusters from the EGRHS, namely,
RXJ0439.0+0520 and Z348, were identified as cool core clusters using the three
criteria. A mini-halo is detected in RXJ1532.9+3201 (Sec. 4.2).
The cluster A689 from the EGRHS sample is also a cool core
cluster. However, as a reliable upper limit on radio halo detection could not be
obtained due to a bright central source, it was
excluded, also from the analysis of mini-halo upper limits.

The injection model profile that we used, is the same as that for radio halos 
but scaled to a maximum size of 500 kpc. The observed profiles
of some mini-halos are
similar to those of radio halos while some are more centrally peaked
 \citep{mur09}. Therefore the model profile
adopted here is not the most accurate choice. However, it is useful to
put conservative upper
limits since the scaled structure of radio halos is more difficult to detect
than possible mini-halos. The model profile was injected in the 610 MHz {\it
uv-}data of each of the five clusters and upper limits were obtained (Table 3). 
The cluster A2667 was excluded due to the presence of a
radio galaxy near its center. The newly obtained upper limits and 
 the known mini-halos, namely, RXCJ1504.1-0248 \citep{gia11}, A1835
\citep{gov09} , Z7160 (V08), RXJ1532.9+3021 and 
A2390 \citep{bac03} from the GRHS sample and Ophiucus, A2029 \citep{gov09},
Perseus \citep{sij93}, RXCJ1347.5-1145 \citep{git07} and 
RBS797 \citep{git06} from the literature are plotted in the
$P_{1.4\mathrm{GHz}}$ - $L_{X}$ plane (Fig.~\ref{lxlrmin}, left). 
 The mini-halos in A2626 \citep{git04}, A2142 
\citep{gio00} and MRC0116+111 \citep{bag09} listed in literature 
\citep[e.g.][]{fer12} are not included \footnote{A2626: nature of diffuse 
emission is uncertain (S. Giacintucci private communication); A2142: there is 
evidence for Mpc-scale 
radio halo \citep[][Farnsworth et al. submitted]{ros13}; MRC0116+111: an upper 
limit on the X-ray luminosity is available.} in the plot.

\section{Discussion}
This paper presents the EGRHS sample and the results from the analysis of 
radio data on 12 clusters. Sensitive radio images of the cluster
fields at 610/235/325 MHz were obtained (Fig.~\ref{clus} and Appendix 1). 
Cluster scale diffuse emission such as radio halo, relic or mini-halo are not
detected in 11 of these clusters. We used the method of radio halo model
profile injection to estimate upper limits on the detections of radio halos
in these clusters. This method was also applied to estimate upper limits on the
detections of mini-halos in a sub-sample of cool-core clusters extracted from
the GRHS and the EGRHS samples.

The mini-halo in the cluster RXJ1532.9+3021 is detected at 610 MHz. A small
scale relic is suspected in Z348, but requires further confirmation. We also
reported instances of detections of diffuse emission associated with
individual galaxies in two of the cluster fields. Synchrotron emission from a
disk galaxy and a radio galaxy with faint lobes were detected in the image of
RXJ0439.0+0715. In the field of A689, a blend of radio galaxies and a patch of
diffuse emission were detected.

We discuss the implications of these results to the understanding of the radio
halos, relics and mini-halos in galaxy clusters.

\subsection{Fractions of radio halos and mini-halos}
The major goal of the GRHS+EGRHS surveys is to estimate the occurrence of
diffuse radio emission in galaxy clusters. We revise the statistics presented in
V08
based on the information on 35 galaxy clusters available then. Out of the full
sample of 67
clusters (GRHS+EGRHS), we now have information (presence of radio halo or upper
limit)
on 48 clusters based on the analysis of our GMRT data, archival (VLA or GMRT)
data and the literature information. Among these 48 clusters, 11 are host to
 giant radio halos\footnote{A2744 \citep{gov01}, A209 (V07), A521 
\citep{bru08}, A1300 \citep{rei99}, A2163 \citep{her94,fer01}, 
RXCJ$2003.5-2323$ (V07), A520 \citep{gov01}, A697 (V08), A773 \citep{gov01}, 
A1758a \citep{gio06}, A2219 \citep{bac03} are the 11 giant radio 
halos. The nature of the radio halo in RXCJ$1314.4-2515$ 
\citep[][V07]{fer05} is uncertain.}.
The fraction of radio halos is thus, $f_{RH} = 11/48 = 23 \pm 7
\%$ \footnote{Error is estimated assuming Poisson statistics.}. Following V08 
and \citet{bru09}, we divided the sample in two X-ray 
luminosity bins, namely, low luminosity
bin, $5\times10^{44}$ erg s$^{-1} < L_X < 8 \times 10^{44}$ erg s$^{-1}$ and
high luminosity bin, $L_X > 8 \times 10^{44}$ erg s$^{-1}$ (see Fig.~\ref{lxlr},
 vertical dashed line). 29 clusters out of
the total 48 belong to the high luminosity bin. Of the 11 radio halos, 
 9 are
hosted in the high luminosity clusters. Thus, the fraction of radio halos in
high luminosity clusters is  $f_{RH} = 9/29 = 31 \pm 10\%$. This fraction 
is consistent with $38 \pm 13\%$, obtained from the sub-sample of 35
clusters in V08. The fraction of radio halos in the lower luminosity bin is, 
$f_{RH} = 2/19 = 11 \pm7\%$.  These fractions refer to the GRHS+EGRHS 
samples that are selected based on X-ray luminosity. However, the X-ray 
lumonisity depends on factors such as the dynamical properties of the cluster 
and the cluster mass. We point out that the derived occurrence of 
radio halos may change when the occurrence with respect to the cluster 
temperature and/or mass is considered (e.g. Cassano et al. in prep.).

We attempt to estimate the mini-halo fraction in the extracted sub-sample of
cool-core clusters. A total of 5 mini-halos are detected so far in the 
GRHS+EGRHS samples and 5 upper limits are obtained. The fraction of mini-halos 
is $\sim 50\%$ in cool-core clusters, indicating their common occurrence in 
cool-cores.

\subsection{Non-detections and cluster dynamics}
The lack of the detections of radio halo/relics in the 11 clusters underlines
the rarity of these sources. The clusters with known radio halos
and relics provide strong evidence that
dynamical disturbances in the clusters play an important role in the generation
of radio halos and relics. Based on the GRHS
sample it has been found that radio halos occur in clusters that show higher
levels of disturbance than the clusters without radio halos \citep{cas10}. 
We searched the literature and carried out a visual examination of the X-ray
images of these clusters to find out about their dynamical states. 
The visual examination of the X-ray images shows that most of these clusters
have no major signatures of disturbances (Fig.~\ref{clus}). Circularly symmetric
or
elliptical morphologies with strong or weak central peak
are seen in the X-ray images of A689, RXJ0439.0+0520, RXJ0439.0+0715, A267,
Z348, A2261 and 
RXCJ0437.1+0043. The clusters A1576, RXJ0142.0+2131,
RXCJ1212.3-1816 and A2485 show relatively disturbed morphologies in X-rays, 
however have low X-ray luminosities ($L_X < 10^{45}$
erg s$^{-1}$). These properties are similar to those of other merging
systems with low X-ray luminosities that have been found to be radio-quiet
(without radio halos/relics) \citep{cas10,rus12}.
The only two clusters in this sub-sample with $L_X>10^{45}$
erg s$^{-1}$, namely A689 and A2261, qualify to be cool cores based on their
properties. Quantitative studies of the dynamics in these clusters will be
pursued 
to understand the non-detections of diffuse radio emission in
these clusters.

The ICM can also be disturbed by the AGN in the cluster centers via feedback
mechanism \citep[see][for a review]{mcn07}. The central AGN during its
cycles of activity produces jets and lobes that can excavate `cavities' in the
ICM
\citep[e.g.][]{clar97,mcn01}.  From the radio images, it is seen that 
 7 clusters (out of the 11), namely, A1576,
RXJ0439.0+0520, RXJ0439.0+0715,
RXJ0142.0+2131, A689, A2261 and Z348, have a strong central radio source
(Fig.~\ref{clus}). The central source in A1576 shows indications of a core and
jets. In other clusters higher resolution observations are required to find
whether the central sources have structure. On a simple visual examination of
the X-ray images, we do not find the presence of obvious cavities in these
clusters. This is in line with the fact that the central sources are compact or
confined to the galactic scales. 
Three of these clusters with central radio sources, namely, RXJ0439.0+0520, Z348
and A689, are confirmed cool core clusters. On the other hand, the clusters
A267, RXCJ0437.1+0043, RXCJ1212.3-1816 and A2485, that have very weak/ no
central radio sources are not cool core clusters. These observations support
 the idea that cool cores are more likely to have a central AGN active in
radio \citep[e.g.][]{sun09}. A detailed examination of the X-ray images and the
radio sources in these clusters are required to obtain further information about
the state of their ICM.

\subsection{The $P_{1.4\mathrm{GHz}}$ - $L_{X}$ correlation}
A scaling between the radio power of radio halos and the X-ray
luminosities
of the clusters is known \citep{lia00, cas06, cas07}. The upper limits
obtained in
this paper for the 10 clusters are factors of 3 to 20 below the expected radio
power based on the correlation (Fig.~\ref{lxlr}). 
The bimodality in the distribution of the clusters in this plane suggests that
radio halos are transient sources connected with merging clusters
\citep[e.g.][]{bru09,cas10}.
 
The bimodality in the $P_{1.4\mathrm{GHz}}$ - $L_{X}$ plane is an important
constraint for the theoretical models. In the turbulent re-acceleration model,
the MHD turbulence generated
in the ICM due to mergers is responsible for the acceleration of particles
\citep[e.g.][]{bru09}. The lifetime of a radio halo depends strongly
on the level of turbulence in the cluster \citep[e.g.][]{cas05, bru09}.
Therefore when the strength of turbulence falls,
the
radio halo fades rapidly. The bimodality in the $P_{1.4\mathrm{GHz}} - L_{X}$
plane can be explained as an outcome of this phenomenon. The key predictions
and observables have been recently reproduced in a high resolution MHD
simulation of the re-acceleration of cosmic ray electrons by turbulence in
cluster mergers \citep{don13}. 
In the case of
secondary electron models the presence of cosmic ray protons is expected in all
clusters and no abrupt change in the level of injection of relativistic
electrons can be expected \citep[e.g.][]{min01, dol00}. Bimodality has been
qualitatively accounted for in
secondary electron models by assuming amplification of magnetic fields in
merging clusters as opposed to relaxed clusters \citep{kus09,kes10}. However,
this is disfavoured by rotation measure studies of galaxy clusters
\citep[][and references therein]{bon11}. There are also recent attempts to
reconcile bimodality and secondary models based on the assumptions that cosmic
ray protons stream at super-Alfvenic speeds in relaxed systems 
\citep{ens11}. 

 The occurrence of USSRHs is one of the expectations of the turbulent 
re-acceleration model \citep[e.g.][]{cas06}. The three USSRHs are found 
to occupy the region between the upper limits and the correlation in the 
$P_{1.4\mathrm{GHz}}$ - $L_{X}$ plane (Fig.~\ref{lxlr}). We note here that the 
upper limits at 1.4 GHz have been scaled from those at 610 MHz, assuming a 
spectral index of 1.3. If a steeper spectral index such as 1.5 or more is used 
then the upper limits will be factors of 1.5 or more deeper. Therefore the 
relative position of the USSRHs and the upper limits needs to be interpreted 
with caution. The dual frequency data opens up a possibility of the detection 
of USSRHs. With the sensitivities that are achieved at 235 and 610 MHz, we can 
infer that radio halos of spectral indices 1.8 or steeper would be detectable 
with our 235 MHz observations.
Systematic radio surveys such as the EGRHS are necessary in
order to populate the $P_{1.4\mathrm{GHz}} - L_{X}$ plane and provide
constraints to the theoretical models. In addition to the 
surveys, characterisation of the
spectra of the radio halos needs to be carried out using
sensitive multi-wavelength observations \citep[e.g.][]{ven13}.

\begin{figure*}
   \centering
    \includegraphics[height=8.5 cm]{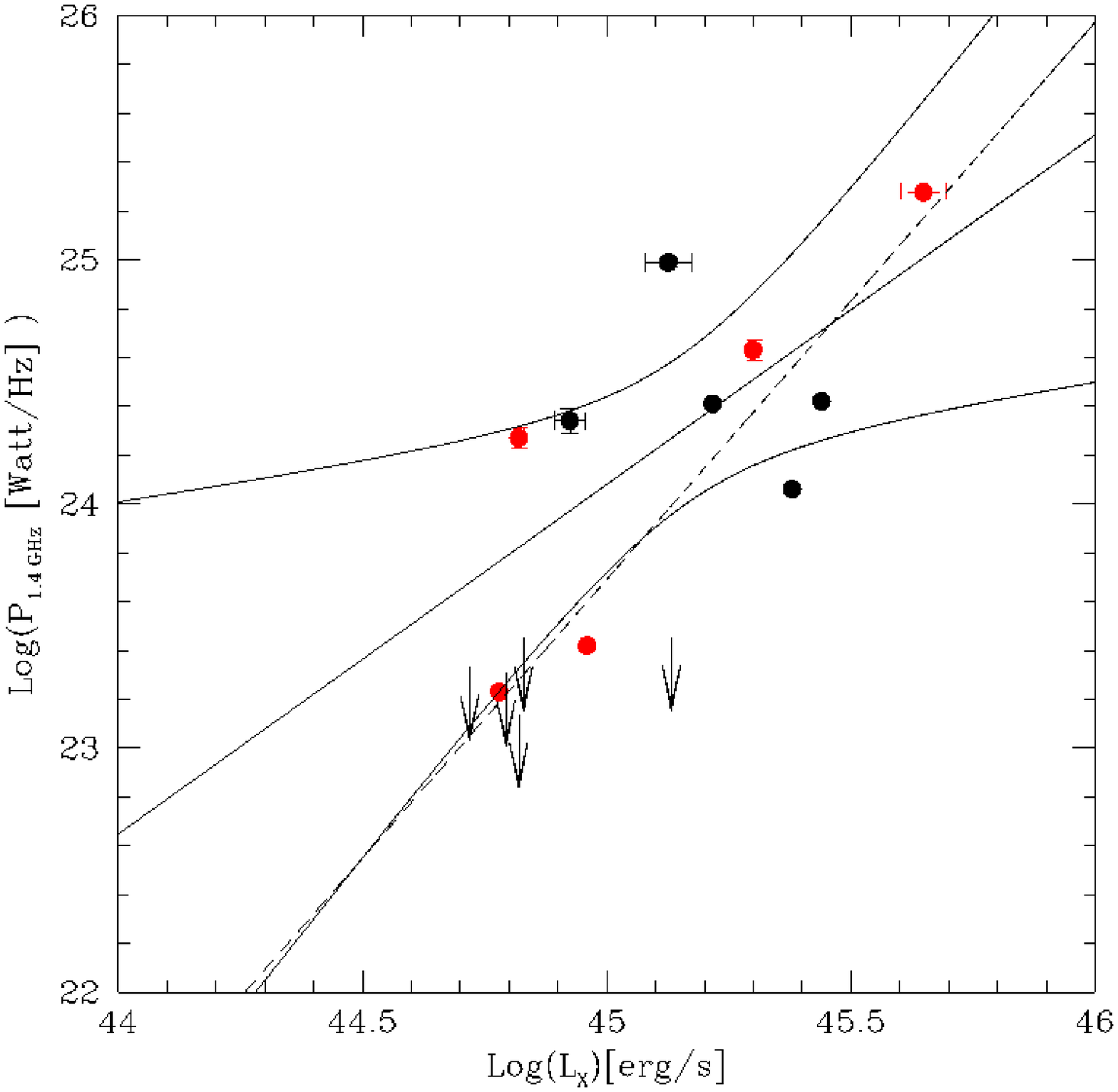}
    \includegraphics[height=8.5 cm]{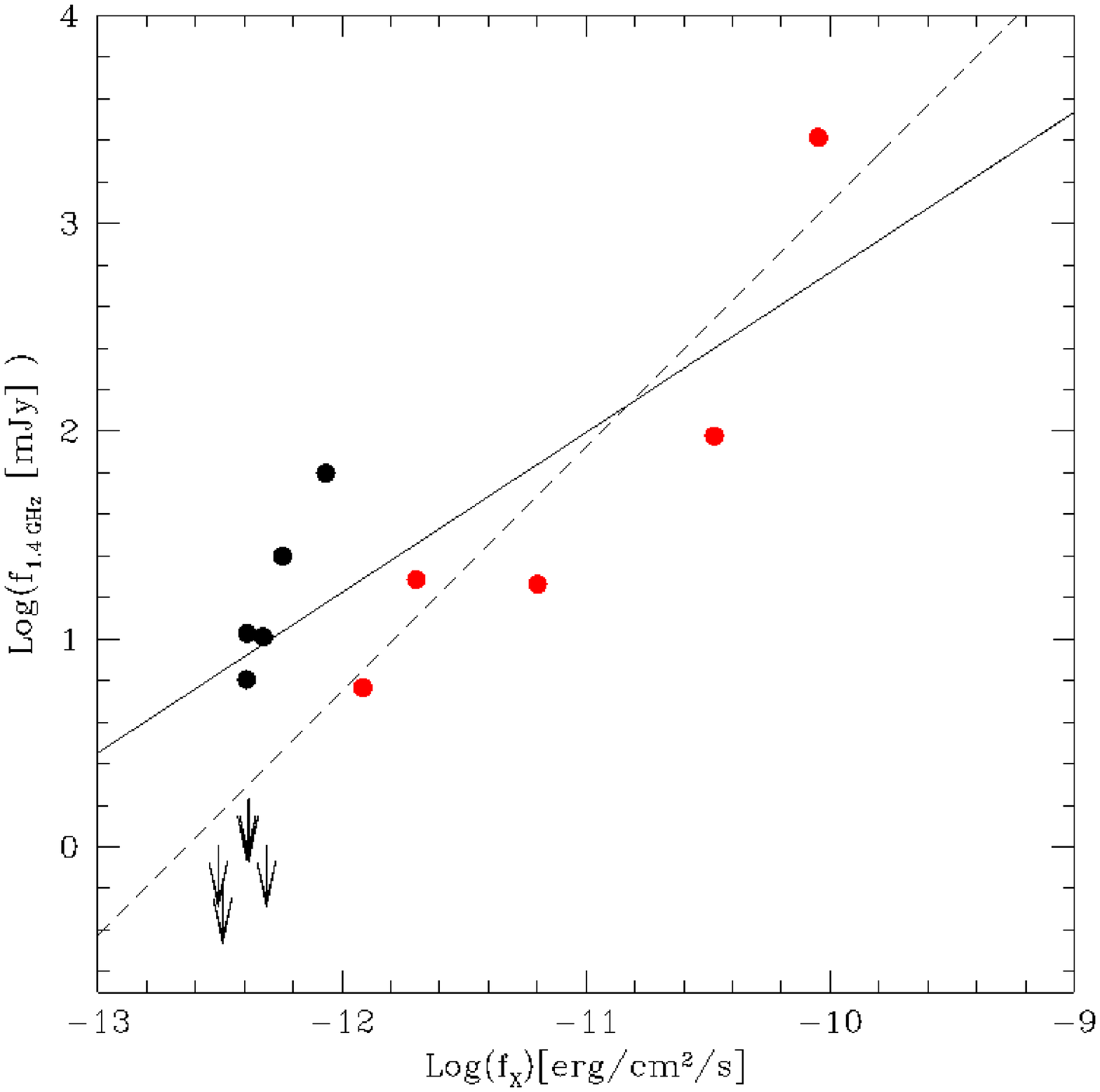}
\caption[]{ 
{\it Left} The $P_{1.4\mathrm{GHz}}$ - $L_{X}$ plane for
mini-halos. The
black arrows are the upper limits on the detection of mini-halos (see Sec.
5.1). The black
filled circles are mini-halos detected in the GRHS sample. Red points are other
mini-halos from the literature \citep{fer12}. The black, solid straight line
is the best fit obtained for a possible
scaling between the
radio-power and X-ray luminosity using only detected mini-halos. The
black curved lines encompass the $95\%$ confidence region (i.e. the
region that has $95\%$ probability to contain the regression line). 
The dashed line is a best fit also including the upper limits. {\it Right} The
distribution of the same mini-halos and upper limits in the radio versus X-ray
flux plane. The solid line has a slope $0.77\pm0.17$ and the dashed line has
a slope $1.17\pm0.25$.}
         \label{lxlrmin}
\end{figure*}

Radio mini halos and their occurrence in galaxy clusters are less explored.
Based on the cool-core sub-sample of clusters from the GRHS and EGRHS, we
find that $\sim 50\%$ of cool core clusters have a mini-halo.
A possible scaling between the radio power of mini-halo
and the X-ray luminosity was reported by \citet{cas08} based on 6 mini-halos
known then. The
combination of GRHS and the EGRHS offers a larger sample to study the statistics
of mini-halos. We used literature data on mini-halos and the upper limits
estimated in this work to examine the scaling between the radio power of
mini-halos and the X-ray luminosity of the cluster (Fig.~\ref{lxlrmin}, left). 

The best fit line using only mini-halo detections has a slope of $1.43\pm0.52$
(solid line).
With the inclusion of upper limits, the best fit has a slope of $2.28\pm0.61$
(dashed line)\footnote{This was estimated through the parametric EM algorithm
as implemented in the ASURV package \citep{iso86}.}. Since the
$P_{1.4\mathrm{GHz}} - L_{X}$ correlation is
consistent with a slope $\sim 1$, we test the reliability of this result by
evaluating the presence of a correlation in the flux-flux space
(Fig.~\ref{lxlrmin}, right). We find a clearer correlation in the flux-flux
plane
with a
best fit slope of $0.77\pm0.17$ (solid line), where only mini-halo detections
are considered. The presence of an even clear scaling in the flux-flux plane
strengthens the hypothesis that there is an intrinsic scaling between the radio
power of mini-halo and the X-ray luminosity. However, there is a large scatter
in the correlation. It may be explained as a result of the differences in the 
properties within the population of mini-halos. Unlike
the radio halos, mini-halos show a large scatter in their average radio
emissivities (power per unit volume) \citep{mur09} indicating a heterogenous
population. However due to small statistics (only 10 mini-halos), the
properties need to be confirmed by future data.

\section{Summary and conclusions}
The EGRHS is an extensive radio survey of massive 
galaxy clusters with 
$L_{X}(\mathrm{0.1 - 2.4}$ keV$) > 5 \times 10^{44}$ erg s$^{-1} $ 
in the redshift range 0.2-0.4. It is an extension of the GRHS survey (V07,
V08). In this paper we presented the EGRHS sample and
first results based on the radio data analysis. The EGRHS sample consists of 17
galaxy clusters, which combined with the GRHS (V08) makes a sample of 67
galaxy clusters. These clusters are being systematically surveyed in radio band
in order to study the statistical properties of the diffuse radio emission in
galaxy clusters. 
 
\noindent The main results of this paper are summarized below:
\begin{enumerate}
\item Radio images with rms noise in the range 45 - 80 $\mu$Jy beam$^{-1}$ at
610 MHz, 0.25 - 0.40 mJy beam$^{-1}$ at 325 MHz and 0.55 -- 1.8 mJy beam$^{-1}$
at 235 MHz were obtained. Single or dual frequency images of 12 clusters are
presented in this paper.
\item The 610 MHz image of the newly detected mini-halo in the GRHS
cluster RXCJ1532.9+3021 is presented (Giacintucci et al., in prep.). 
 No cluster-scale diffuse emission associated with the remaining 11 clusters
were detected. A small
scale ($\sim200$ kpc) relic is suspected in the cluster Z348.
 The X-ray images of these clusters and the
information in
the literature about the dynamical states were examined. None of these 11
clusters show extreme merger signatures combined with high X-ray luminosities.
\item The method of injection of model radio halos was used to obtain 
firm upper limits on 10 clusters (A1576, RXJ0439.0+0520, RXJ0439.0+0715,
RXJ0142.0+2131, A267, Z348, A2261, RXCJ0437.1+0043, RXCJ1212.3-1816 and A2485);
reliable upper limit for the cluster A689 could not be obtained due to the
presence of a strong source at its center.
\item The new upper limits are plotted in the $P_{1.4\mathrm{GHz}}$ -- $L_X$
plane and are found to be factors $\sim$ 3 - 20 times below the expected radio
power
based on the correlation.
The fraction of radio halos in the EGRHS+GRHS sample analysed so far is
{\bf $23 \pm 7 \%$}. This fraction is {\bf $31 \pm 11\%$} in the clusters with 
high X-ray
luminosities ($>8\times10^{44}$ erg s$^{-1}$) and $11\pm7\%$ in the clusters
with lower X-ray luminosities ($5-8 \times10^{44}$ erg s$^{-1}$).
\item From the GRHS sample and the EGRHS clusters presented here, a sub-sample
of 7 cool core clusters was identified using the criteria based on
central entropy, central
cooling time and luminosity ratio. The method of obtaining upper limits
based on model injection is extended to mini-halos. Upper
limits on the detection of mini-halos were obtained for 5 cool core clusters
(RXCJ1115.8+0129, Z2089, Z2701, RXJ0439.0+0520 and Z348). The plot of
$P_{1.4\mathrm{GHz}}$ -- $L_X$ for mini-halos, including the 
data from literature, is presented. There is an indication of a correlation
which needs to be confirmed by future data. In the GRHS+EGRHS cool-core
sub-sample, the fraction of mini-halos 
is found to be $\sim50\%$.
\item Detections of low brightness, extended radio emission, not related to the
clusters themselves but detected in the sensitive radio images presented here
are also reported. We have detected emission from a disk galaxy 
(NGC 1633) and possible diffuse lobes of a radio galaxy in the field of the
cluster RXJ0439.0$+$0715. A complex blend of radio galaxies and a diffuse
filament
are detected in the field of A689.
\end{enumerate}
The results on the remaining clusters in the EGRHS
sample and the statistics
of radio halos and mini-halos based on the full sample will be presented in a
future work.

\begin{acknowledgements}
We thank the staff of the GMRT who have made these observations possible. 
GMRT is run by the National Centre for Radio Astrophysics of the Tata Institute
of Fundamental Research. This research has made use of the NASA/IPAC
Extragalactic Database (NED) which is operated by the Jet Propulsion Laboratory,
California Institute of Technology, under contract with the National Aeronautics
and Space Administration. We have made use of the ROSAT Data Archive of the
Max-Planck-Institut fur extraterrestrische Physik (MPE) at Garching, Germany.
This research has made use of data obtained from the High Energy Astrophysics
Science Archive Research Center (HEASARC), provided by NASA's Goddard Space
Flight Center. SG acknowledges the support of NASA through Einstein Postdoctoral
Fellowship PF0-110071 awarded by the Chandra
X-ray Center (CXC), which is operated by SAO. 
 This work is partially supported by PRIN-INAF2008 and by
FP-7-PEOPLE-2009-IRSES CAFEGroups project grant agreement 247653. GM 
acknowledges financial support by the ``Agence Nationale de la 
Recherche'' through grant ANR-09-JCJC-0001-01.
\end{acknowledgements}

\bibliographystyle{aa}
\bibliography{a3376}

\begin{table*}[]
\caption[]{\label{t1}Extended GMRT RHS Sample (first sector, 17 clusters) and
part of
the GRHS sample.
Columns are: 1. Cluster name; 2. Right Ascension; 3. Declination; 4. Redshift;
5. X-ray
luminosity$^1$ 
(0.1 -- 2 keV) in units $10^{44}$  erg s$^{-1}$ \citep{ebe98,ebe00,boh04}; 6. 
Notes about the cluster. Radio data on the clusters marked with the symbol 
`$\surd$' are presented in this paper.}
\begin{center}
\begin{tabular}{llccrc}
\hline
Name&  RA$_{J2000}$ & DEC$_{J2000}$ & z & L$_X$ (0.1--2 keV)     &
Notes \\
                &              &               &   & $10^{44}$ erg
s$^{-1}$&        \\
\hline
A68	                 &  00 36 59.4 &  +09 08 30 & 0.254 &
9.47  &   \\
Z348                &  01 06 50.3 &  +01 03 17 & 0.254 & 6.23 
& $\surd$\\
RXJ0142.0$+$2131    &  01 42 03.1 &  +21 30 39 & 0.280 & 6.41 
& $\surd$\\
A267	          &  01 52 52.2 &  +01 02 46 & 0.230   & 8.57  & $\surd$\\
RXJ0439.0$+$0715         &  04 39 01.2 &  +07 15 36 & 0.244 &
8.37 &$\surd$\\
RXJ0439.0$+$0520         &  04 39 02.2 & +05 20 43  & 0.208  &
5.30 &$\surd$ \\
 A520 & 04 54 19.0 & +02 56 49 & 0.203 & 8.84 & Radio halo$^2$\\
A689                     &  08 37 29.7 &  +14 59 29 & 0.279 &
19.68 & $\surd$\\
Z1953     &  08 50 10.1 &  +36 05 09 & 0.373 &
23.46 & \\
 Z3146                     &  10 23 39.6 &  +04 11 10 & 0.290 &
17.26 
& \\
Z5247                    &  12 34 17.3 &  +09 46 12 & 0.229  & 6.32
 & \\
A1576   &  12 36 49.1 &  +63 11 30 & 0.302  & 7.20   & $\surd$\\
 A1722                    &  13 19 43.0 &  +70 06 17 & 0.327 &
10.78 & \\
A1835     &  14 01 02.0 &  +02 51 32 & 0.252 & 24.48  & Mini-halo$^3$\\
A2146     &  15 56 04.7 &  +66 20 24 & 0.234 & 5.62   & No halo$^4$\\
 RXJ2129.6$+$0005	  &  21 29 37.9 &  +00 05 39 & 0.235  & 11.66 
& \\
A2552                     &  23 11 33.1 &  +03 38 07 & 0.301 &
10.07 & \\
\hline
 RXCJ0437.1+0043 & 04 37 10.1 & +00 43 38 & 0.284 &8.99&$\surd$ \\
RXCJ1212.3$-$1816        &  12 12 18.9 &  -18 16 43 & 0.269 & 6.20
&$\surd$ \\
A2261               &  17 22 28.3 &  +32 09 13 & 0.224 & 11.31 &$\surd$\\
 A2485 & 22 48 32.9 & -16 06 23 & 0.247 & 5.10 & $\surd$\\
 RXJ1532.9+3021 & 15 32 54.2 & 30 21 11 & 0.345 & 16.48 & $\surd$,
Mini-halo$^5$ \\
\hline
\end{tabular}
\end{center}
\tablefoot{$^1$ Corrected for cosmology; $^2$ \citet{gov01};
$^3$ \citet{mur09};
$^4$ \citet{rus11};
$^5$ Giacintucci et al. 2013, in prep.}
\end{table*}

\begin{center}
\begin{table}[]
 \caption{\label{t2} Summary of GMRT observations.}
\begin{tabular}{llll}
\hline
Cluster&Freq.&Beam & rms \\
 &(MHz)&$''\times''$, PA ($^{\circ}$) & mJy beam$^{-1}$ \\
\hline
Abell 1576 &610& $8.1\times6.7$, $53.2$ & 0.08 \\
           &235& $24.6\times17.8$, $43.8$   &0.55 \\
Abell 689  &610&$7.6\times6.2$, $-63.6$ & 0.08 \\
           &235&$29.7\times25.2$, $35.7$ & 0.85 \\
RXJ0439.0+0520& 610 &$7.8\times6.9$, $-34.0$ &0.05 \\
              &235 & $16.7\times15.8$, $-87.2$ & 1.8\\
RXJ0439.0+0715 & 610&$8.6\times6.8$, $71.1$&0.06
\\
& 235&$27.3\times15.2$, 60.4&0.75
\\
RXJ0142.0+2131 &610 &$9.5\times7.1$, 88.2 &0.05 \\
&235 &$27.3\times13.5$, 54.2 & 1.3\\
Abell 267 &610 &$6.3\times4.4$, 70.6  &0.07  \\
Z348 &610&$14.6\times7.7$, -13.7 &0.065\\
Abell 2261 &610& $11.2\times8.9$, 76.0&0.08 \\
           &235 & $24.4\times11.8$, $-19.0$ & 0.45\\
RXCJ0437.1+0043 &325&$18.6\times14.3$, 82.0  &0.25 \\
RXCJ1212.3-1816 &325&$29.6\times19.7$, 10.4 &0.40  \\
Abell 2485 &325&$18.2\times15.4$, 36.2 & 0.25 \\
RXJ1532.9+3021 & 610 &
$5.6\times5.1$, $82$ & 0.04 \\
\hline
\end{tabular}
\end{table}

\end{center}

\begin{table}[]
\caption{\label{t4} Radio power upper limits using radio halo (first 11) and 
mini-halo injections. See Sec. 5 for details.}
\centering
\begin{tabular}{lclll}
\hline
Cluster& z & rms$^a$&S$_{610\mathrm{MHz}}$ &Log ($P_{1.4\; \mathrm{GHz}}$\\
Name & & mJy b$^{-1}$&mJy& W Hz$^{-1}$ ) \\
\hline\hline
Abell 1576 & 0.302 &0.09&6& 23.78 \\
RXJ0439.0+0520  & 0.208 &0.06&7 & 23.47 \\
RXJ0439.0+0715  & 0.244&0.07&7& 23.63 \\
RXJ0142.0+2131  & 0.280&0.07&5& 23.62 \\
Abell 689$^b$  & 0.279&0.10&-& - \\
Abell 267 & 0.230 &0.15&6& 23.50 \\
Z348 & 0.254&0.13&9& 23.78\\
Abell 2261  & 0.224&0.06&6& 23.48\\
RXCJ0437.1+0043$^c$ & 0.284&0.25 &7& 23.78 \\
RXCJ1212.3-1816$^c$ &0.269 &0.40&7& 23.73 \\
Abell 2485$^c$ & 0.247&0.25& 7& 23.64\\
\hline
\hline
RXCJ1115.8+0129 & 0.350 &0.045& 2 & 23.45\\
Z2089 & 0.234&0.045&5 & 23.44\\
Z2701 &0.214 &0.075 &3&23.13 \\
Z348 & 0.254&0.13 &3 & 23.30\\
RXJ0439.0+0520 & 0.208&0.06 & 5 & 23.33 \\
Abell 2667$^d$ &0.226&0.10 & -&- \\
\hline
\end{tabular}
\tablefoot{$^a$ Rms noise in images tapered to HPBW $\sim 18''-25''$.
$^b$ Reliable upper limit could not be obtained due to a central 
bright source in the cluster. $^c$ The upper limits are scaled from those
at 
325 MHz using a spectral index of $1.3$. $^d$ Upper limit was not obtained
due to a radio galaxy near the cluster centre. }
\end{table}

\Online

\begin{appendix} 
\section{Radio images of the cluster fields}
The radio images of the clusters up to the virial radius are presented here.
Natural weights were used to make these images.

\begin{figure*}
\centering
\includegraphics[width=9cm]{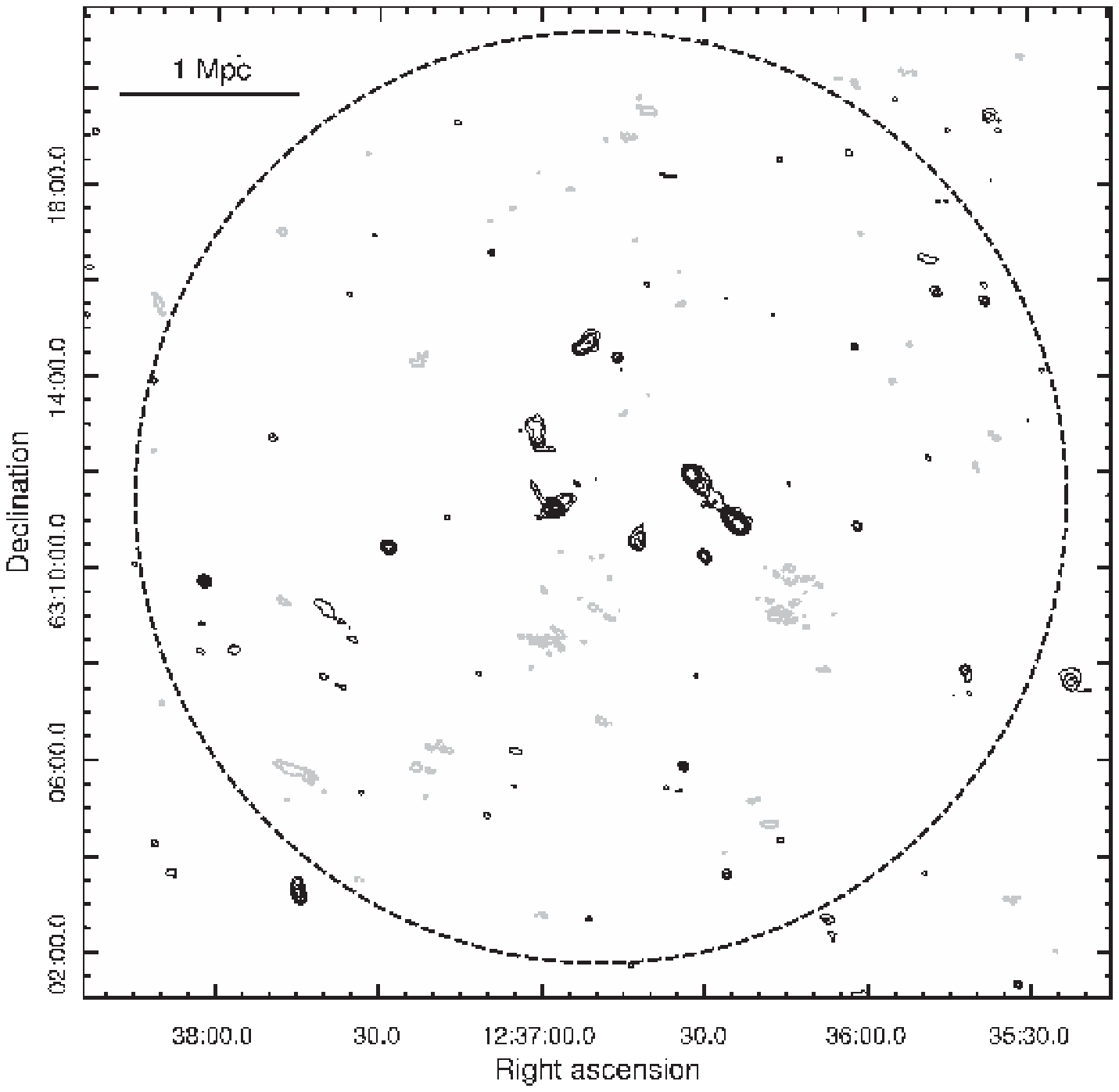}
\includegraphics[width=9cm]{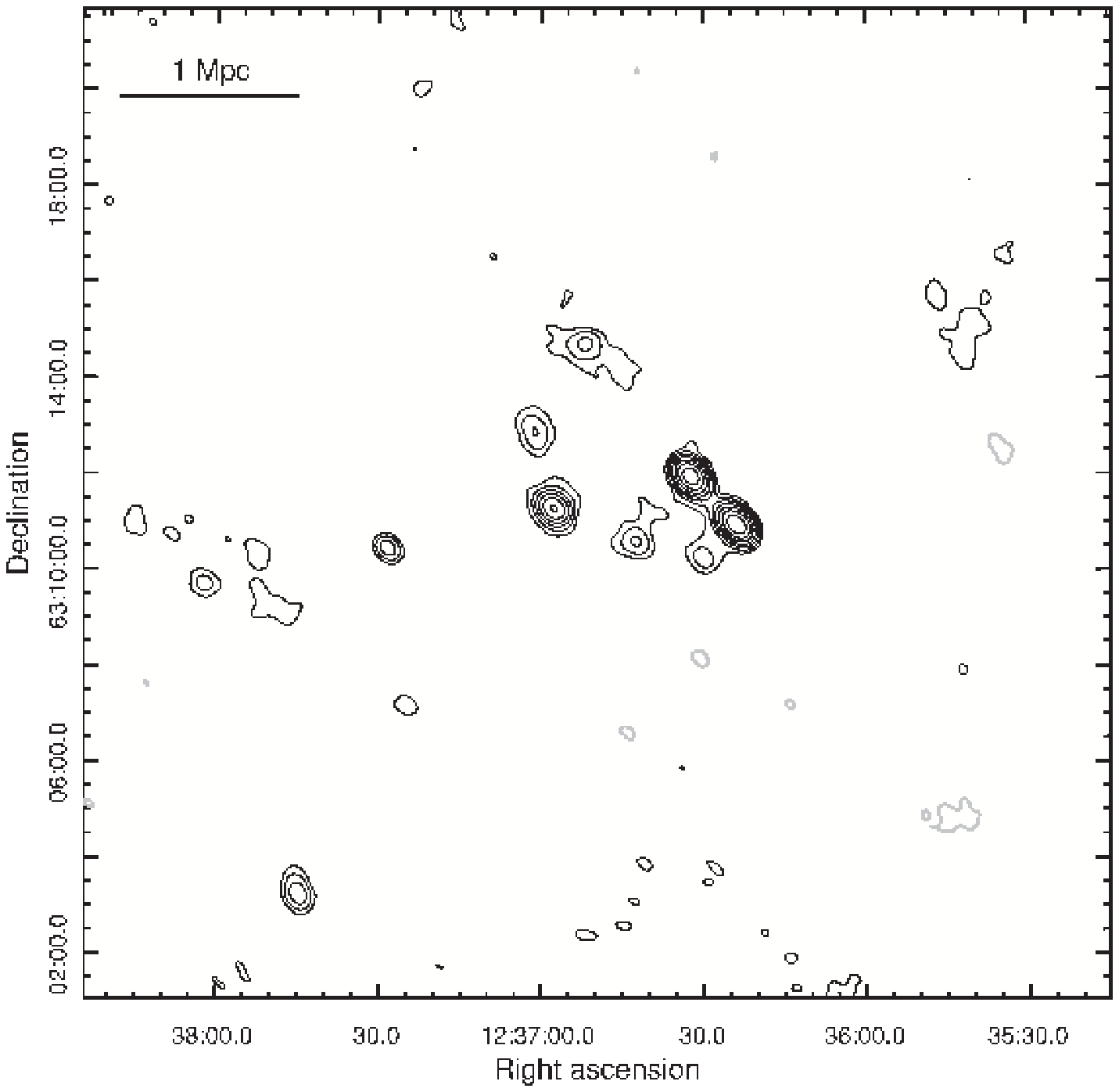}
\caption{{\bf A1576}: GMRT 610 MHz (left) and 235 MHz (right) images in 
contours. Positive
contours are black and negative are grey in all the images in Appendix A. 
The contours are at $0.3\times(\pm1, 2, 4,...)$ mJy beam$^{-1}$ at 610 MHz 
and $2.0\times(\pm1, 2, 4,...)$ mJy beam$^{-1}$ at 235 MHz.  The
HPBWs at 610 and 235 MHz are $8.1''\times6.7''$, p. a. $53.2^{\circ}$ and 
$24.6''\times17.8''$, p. a. $43.8^{\circ}$, respectively. The 
$1\sigma$ levels in the 610 and 235 MHz images are 0.08 and 
0.55 mJy beam$^{-1}$, respectively. The circle has a radius equal to the virial 
radius for this cluster ($9.70'$).}
\label{appa1576}
\end{figure*}


\begin{figure*}
\centering
\includegraphics[width=9cm]{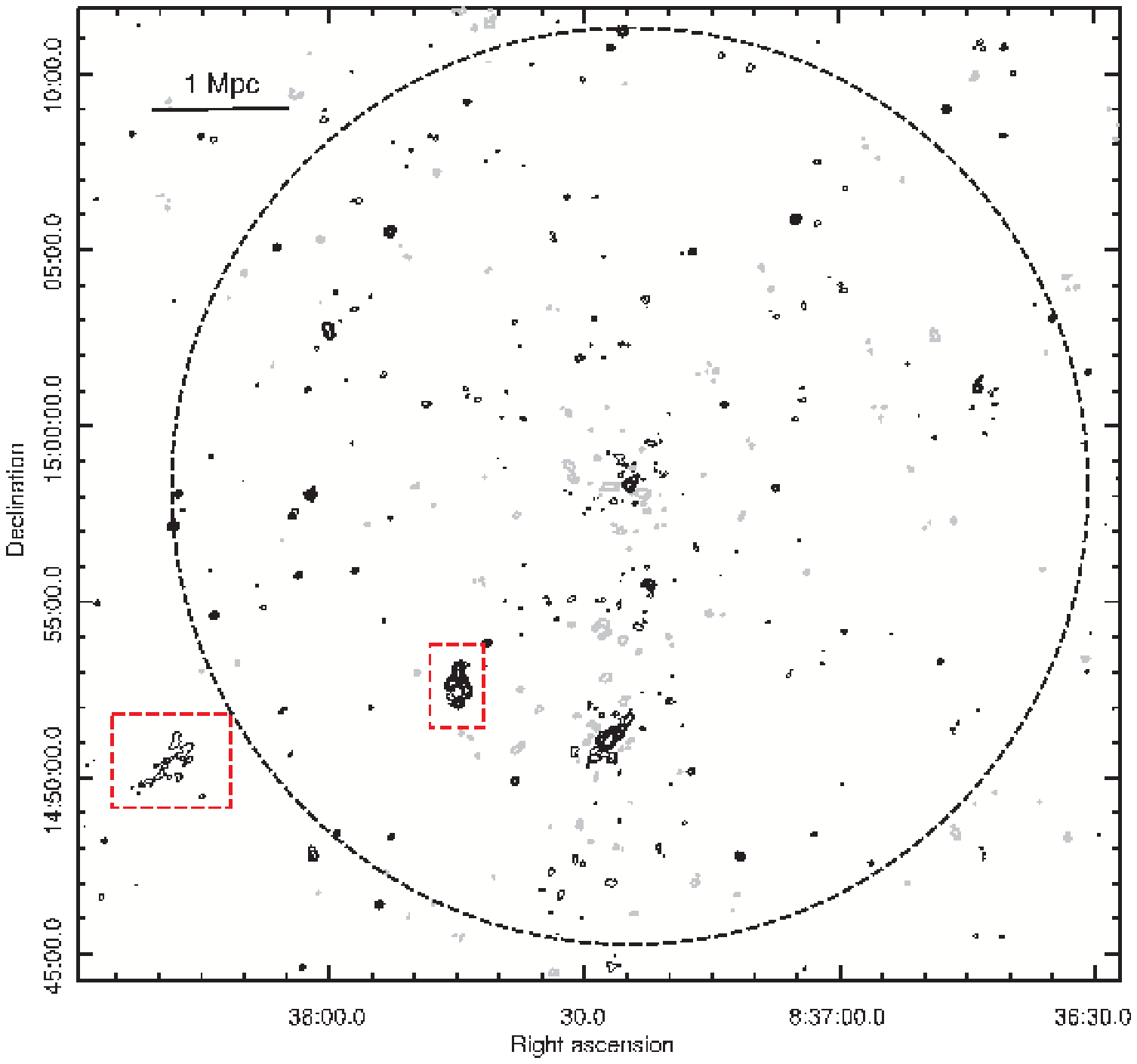}
\includegraphics[width=9cm]{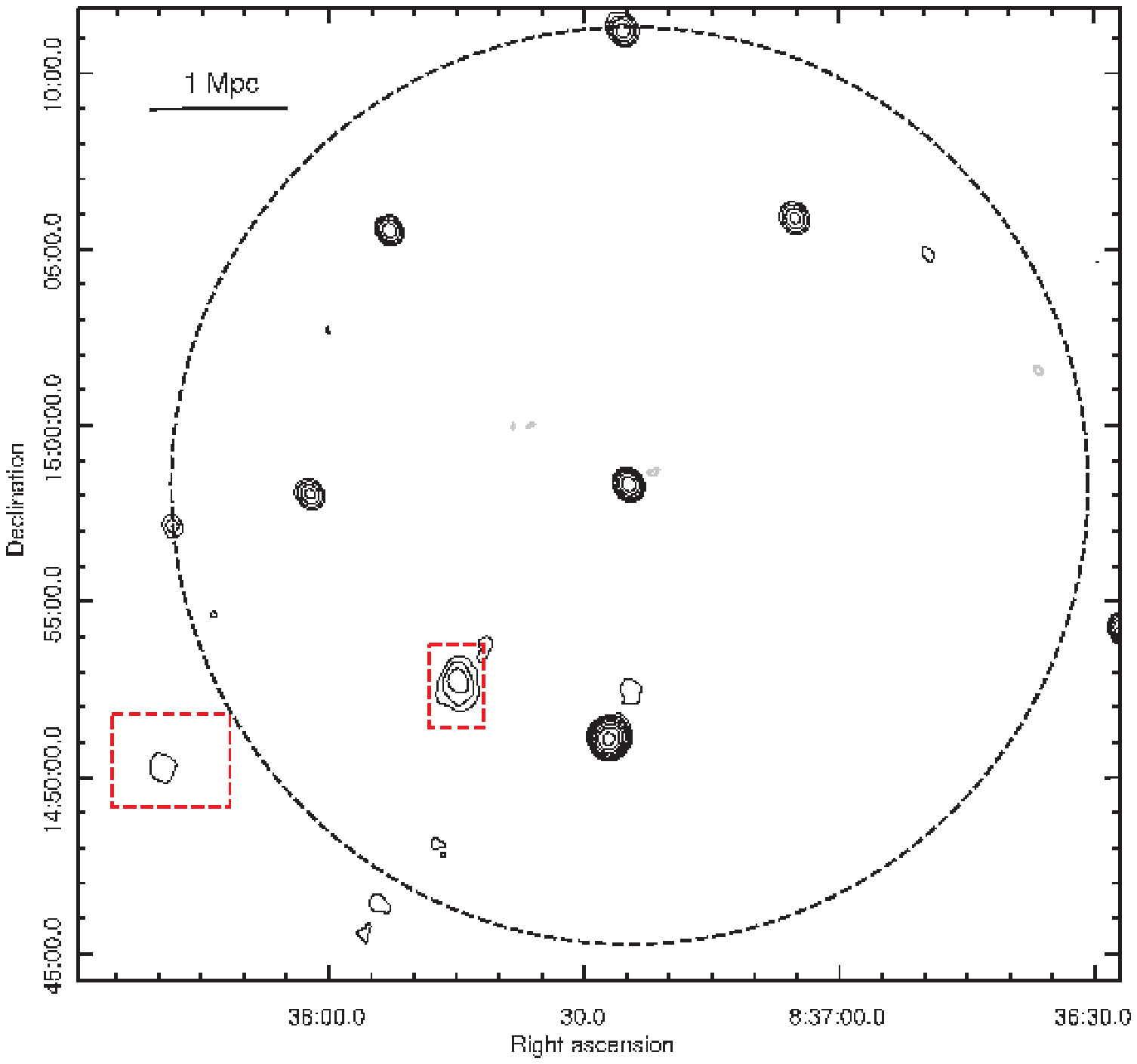}
\caption{{\bf A689}: GMRT 610 MHz (left) and 235 MHz (right) images in 
contours. The contours are at $0.24\times(\pm1, 2, 4,...)$ mJy beam$^{-1}$ 
at 610 MHz and $3.0\times(\pm1, 2, 4,...)$ mJy beam$^{-1}$ at 235 MHz. The
HPBWs at 610 MHz and 235 MHz are $7.6''\times6.2''$, p. a. $-63.6^{\circ}$ and 
$29.7''\times25.2''$, p. a. $35.7^{\circ}$, respectively. The 
$1\sigma$ levels in the 610 and 235 MHz images are 0.08 and 
0.85 mJy beam$^{-1}$, respectively. The circle has a radius equal to the virial 
radius
for this cluster ($13.02'$). The red boxes mark the positions of the diffuse
filament (left box) and the complex of radio galaxies (right box) discussed in
Sec. 4.2.2.}
\label{appa689}
\end{figure*}


\begin{figure*}
\centering
\includegraphics[width=9cm]{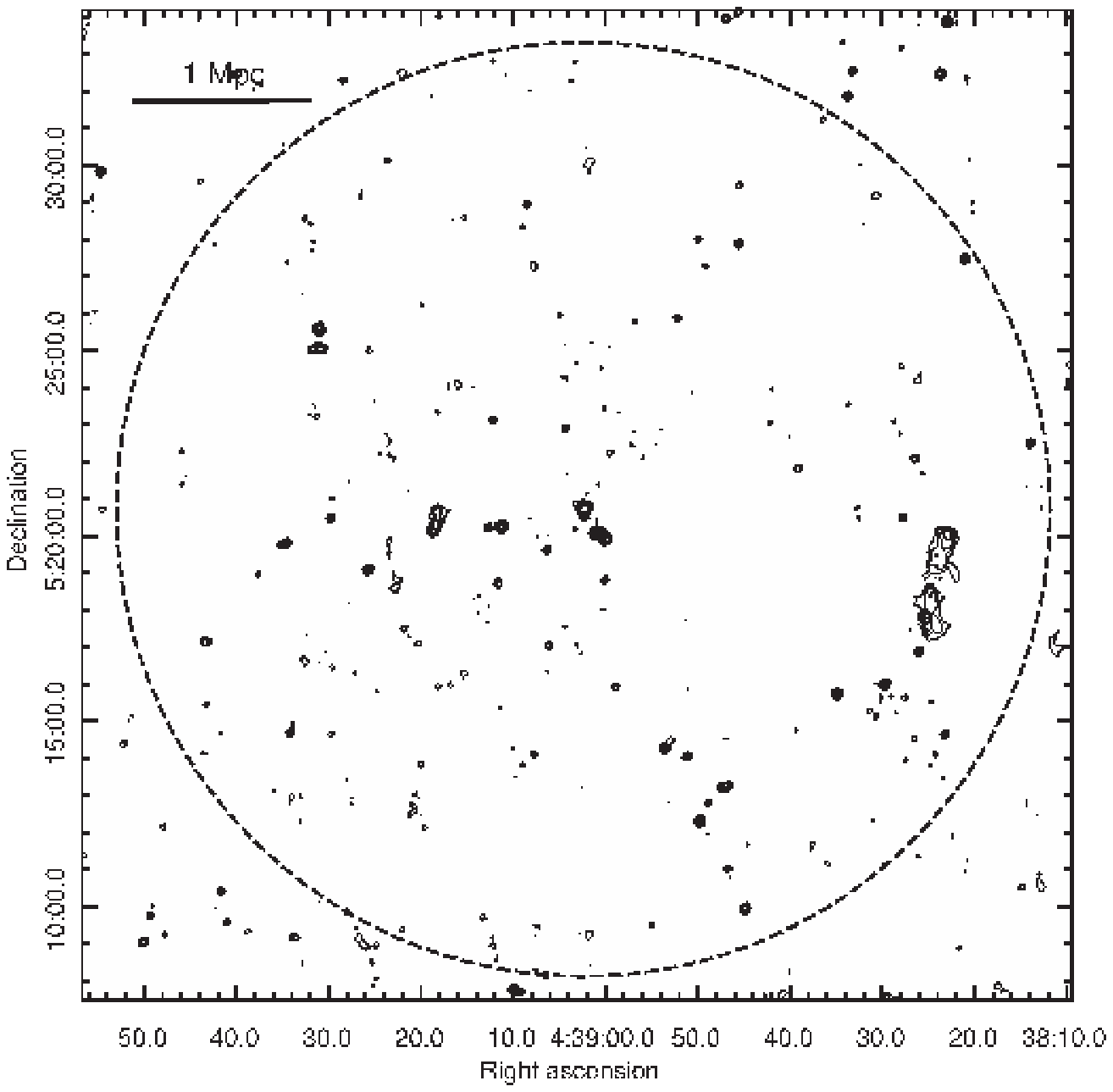}
\includegraphics[width=9cm]{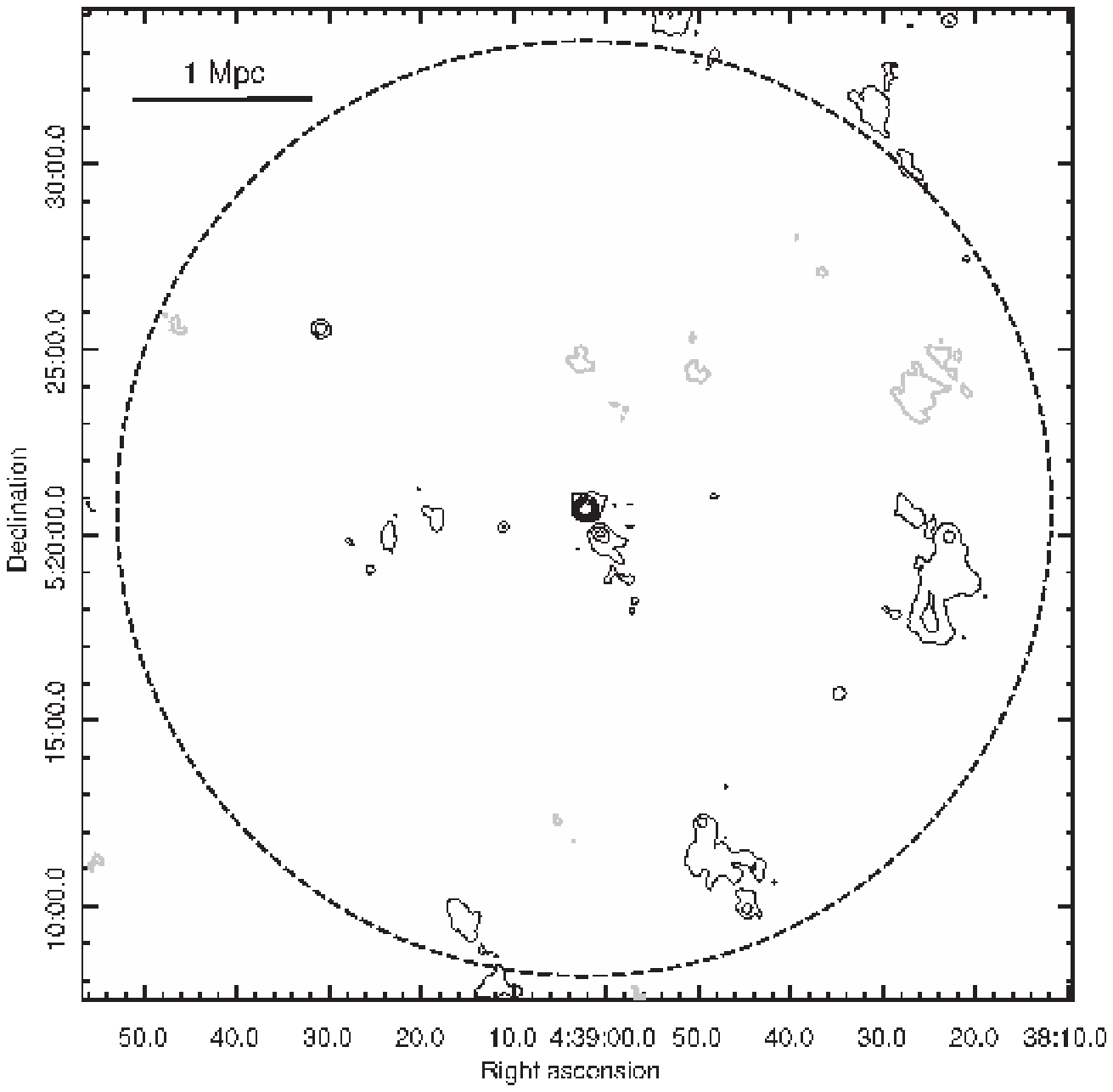}
\caption{{\bf RXJ0439.0+0520}: GMRT 610 MHz (left) and 235 MHz (right) images 
in 
contours. The contours are at $0.15\times(\pm1, 2, 4,...)$ mJy beam$^{-1}$ 
at 
610 MHz and $6.0\times(\pm1, 2, 4,...)$ mJy beam$^{-1}$ at 235 MHz. The
HPBWs at 610 and 235 MHz are $7.8''\times6.9''$, p. a. $-34.0^{\circ}$ and 
$16.7''\times15.8''$, p. a. $-87.2^{\circ}$, respectively. The 
$1\sigma$ levels in the 610 and 235 MHz images are 0.05 and 
1.8 mJy beam$^{-1}$, respectively. The 
circle has a
radius equal to the virial radius
for this cluster ($12.60'$).}
\label{app5rx}
\end{figure*}


\begin{figure*}
\centering
\includegraphics[width=9cm]{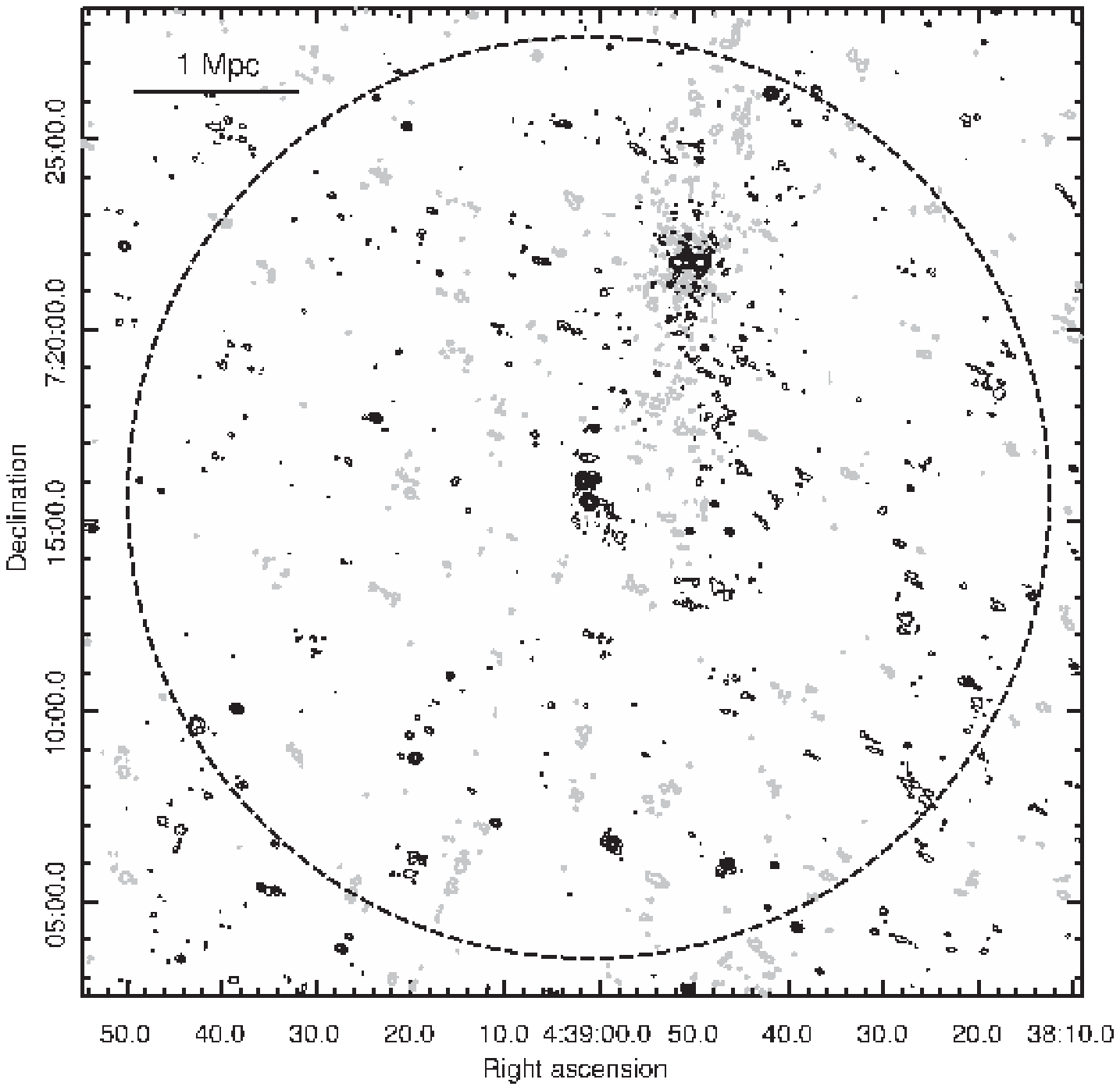}
\includegraphics[width=9cm]{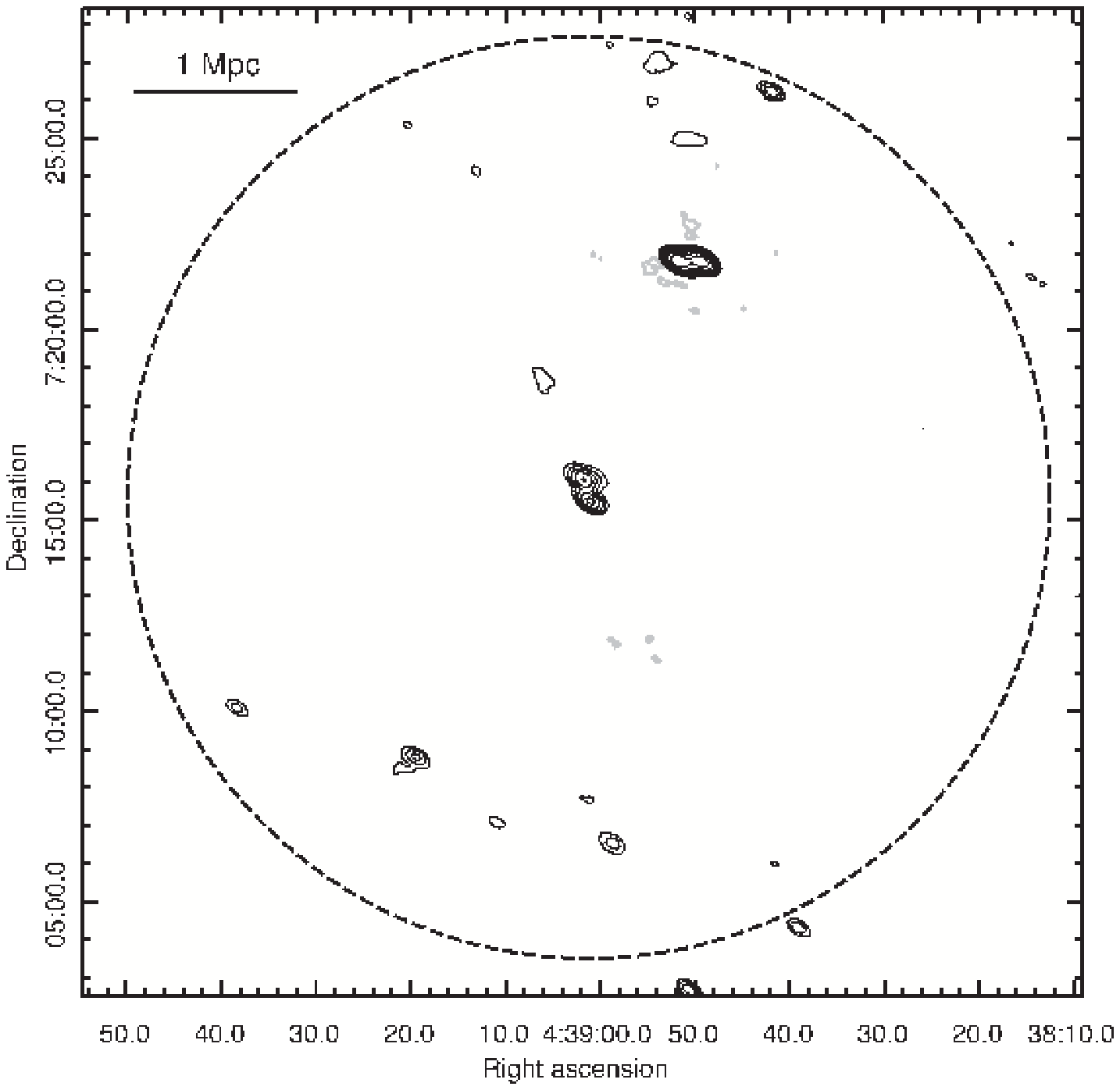}
\caption{{\bf RXJ0439.0+0715}: GMRT 610 MHz (left) and 235 MHz (right) images 
in 
contours. The
contours are at $0.18\times(\pm1, 2, 4, ...)$ mJy beam$^{-1}$ at 610 MHz and 
$3.0\times(\pm1, 2, 4,...)$ mJy beam$^{-1}$ at 235 MHz. 
The HPBWs at 610 and 235 MHz are $8.6''\times6.8''$, p. a. $71.1^{\circ}$ and 
$27.3''\times15.2''$, p. a. $60.4^{\circ}$, respectively. The 
$1\sigma$ levels in the 610 and 235 MHz images are 0.06 and 
0.75 mJy beam$^{-1}$, respectively. The circle has a
radius equal to the virial radius for this cluster ($12.08'$).}
\label{app7rx}
\end{figure*}

\begin{figure*}
\centering
\includegraphics[width=9cm]{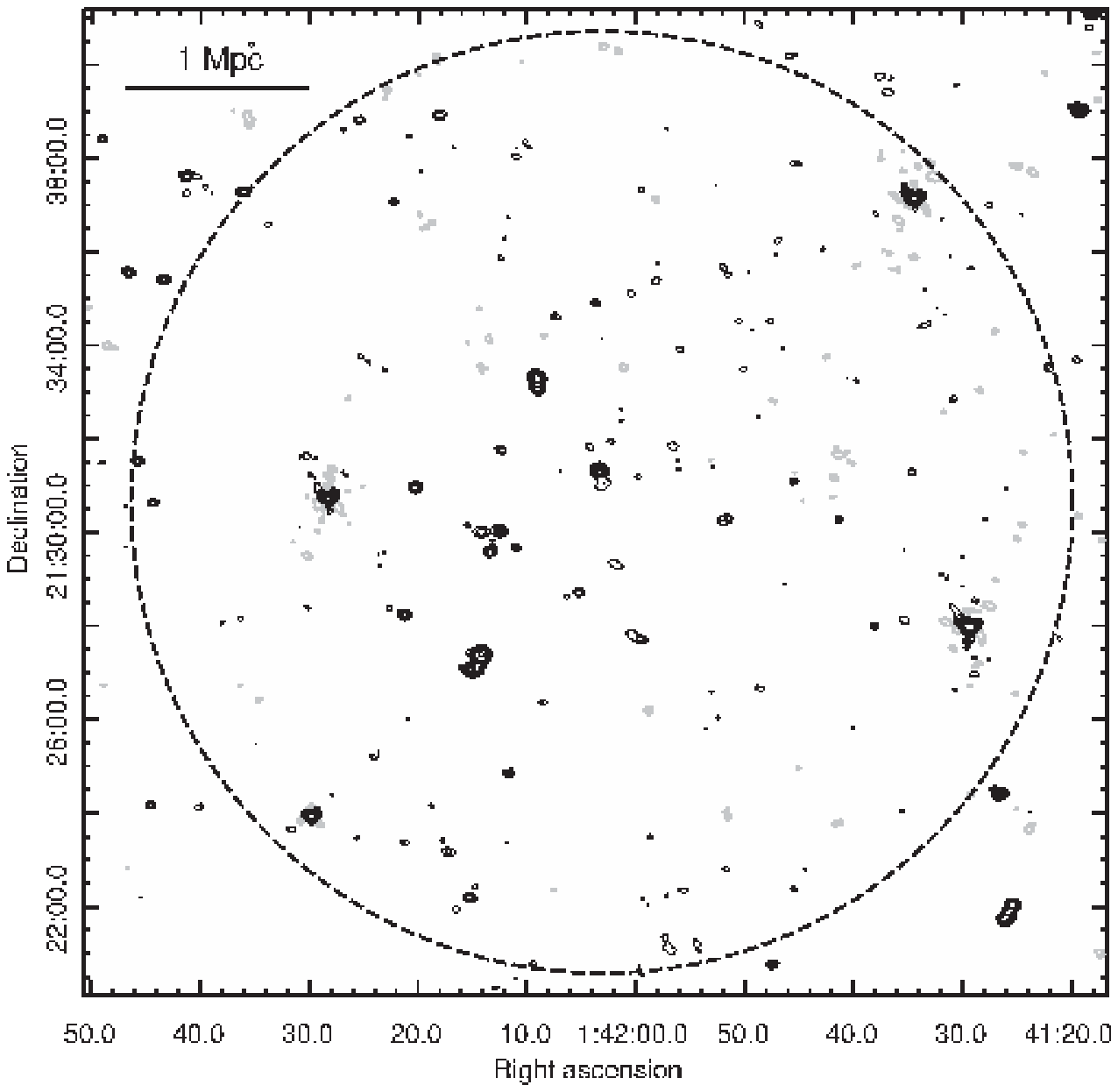}
\includegraphics[width=9cm]{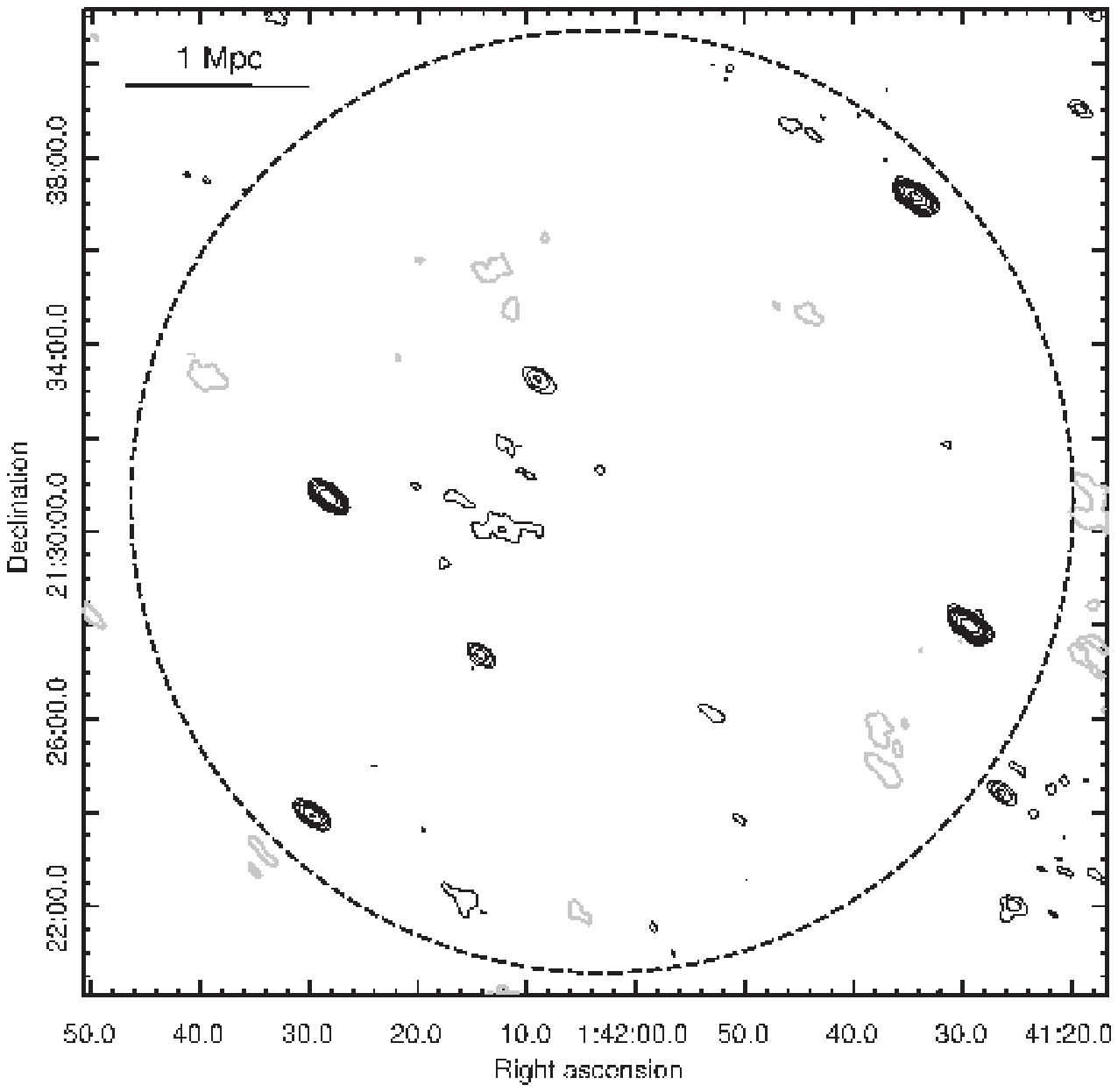}
\caption{{\bf RXJ0142.0+2131}: GMRT 610 MHz (left) and 235 MHz (right) images 
in 
contours. The contours are at $0.2\times(\pm1, 2, 4,...)$ mJy beam$^{-1}$ at 
610 MHz and $4.5\times(\pm1, 2, 4,...)$ mJy beam$^{-1}$ at 235 MHz. The HPBWs 
at 610 and 
235 MHz are $9.5''\times7.1''$, p. a. $88.2^\circ$ and $27.3''\times13.5''$, p. 
a. $54.2^{\circ}$, respectively. The $1\sigma$ levels 
in the 610 and 235 MHz images are 0.05 and 
1.3 mJy beam$^{-1}$, respectively. The circle has a
radius equal to the virial radius
for this cluster ($10.07'$).}
\label{apprx0142}
\end{figure*}

\begin{figure*}
\centering
\includegraphics[width=10cm]{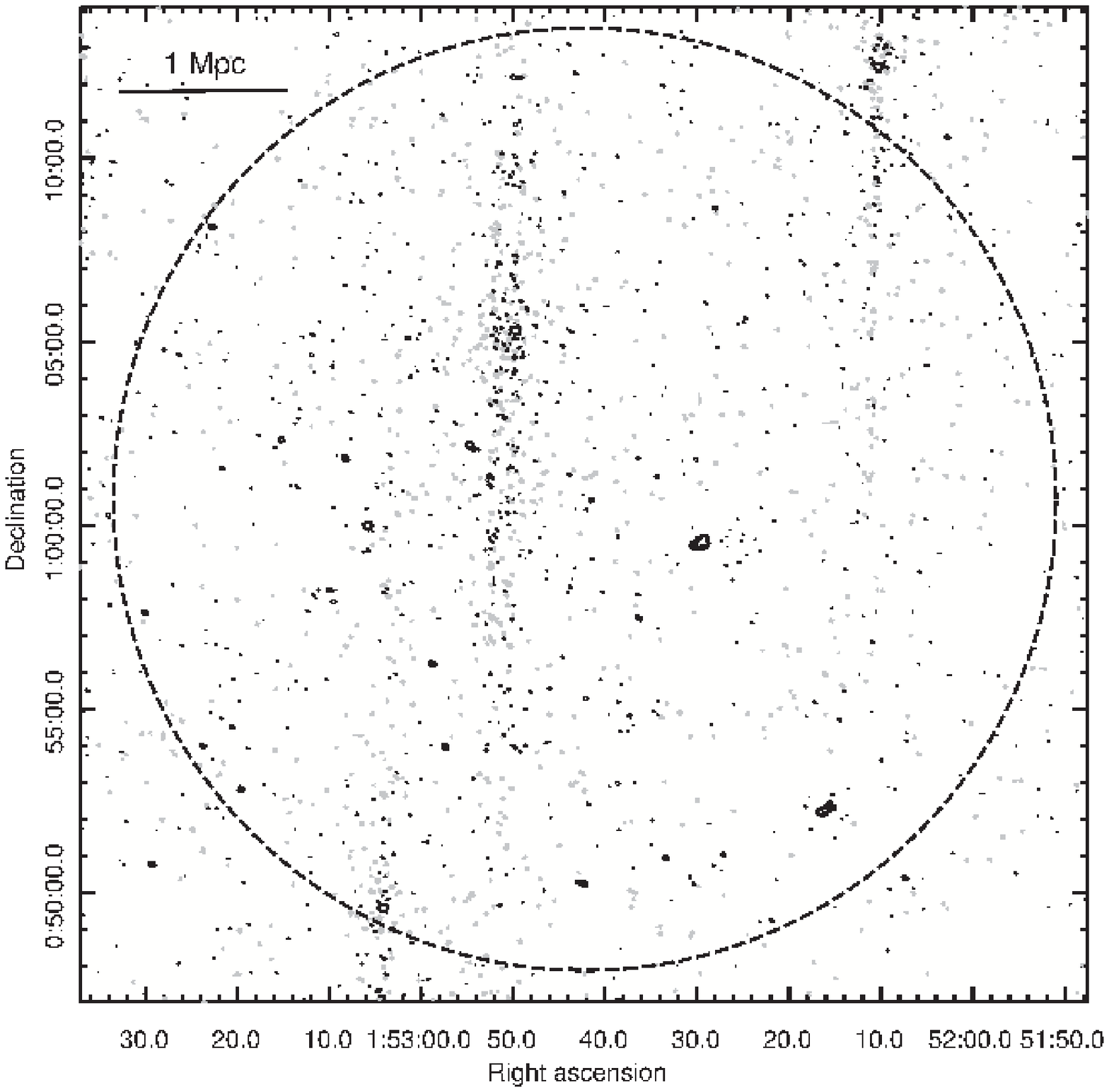}
\caption{{\bf A267}: GMRT 610 MHz image in contours. The
contours are at $0.24\times(\pm1, 2, 4,...)$ mJy beam$^{-1}$. The
HPBW
is $6.3''\times4.4''$, p. a. $70.6^{\circ}$. The 
$1\sigma$ level 
in the image is 0.07 mJy beam$^{-1}$. The 
circle has a
radius equal to the virial radius
for this cluster ($12.82'$).}
\label{appa267}
\end{figure*}

\begin{figure*}
\centering
\includegraphics[width=10cm]{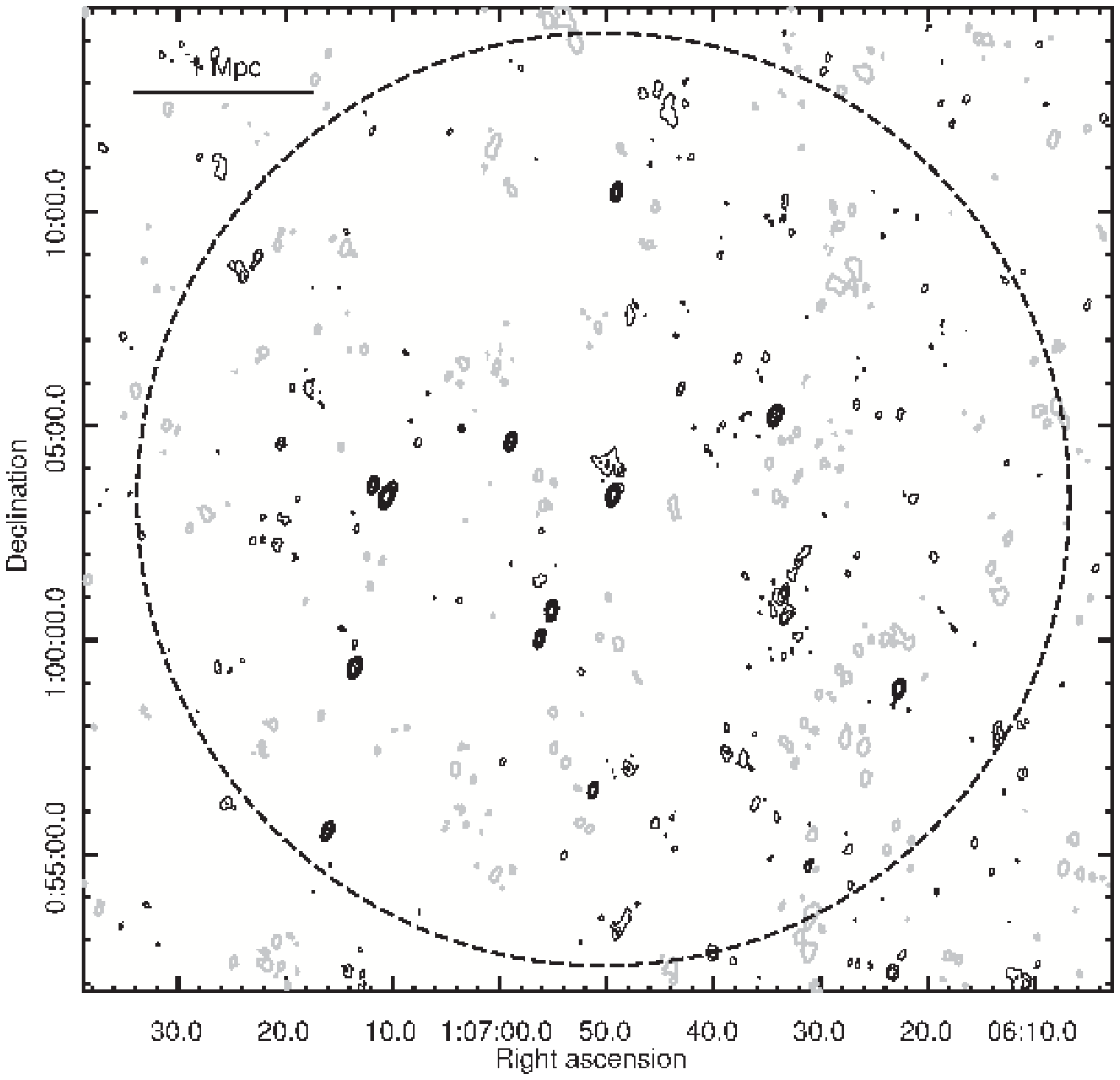}
\caption{{\bf Z348}: GMRT 610 MHz image in contours. The
contours are at $0.2\times(\pm1, 2, 4,...)$ mJy beam$^{-1}$. The HPBW
is $14.6''\times7.7''$, p. a. $-13.7^{\circ}$. 
The 
$1\sigma$ level 
in the image is 0.065 mJy beam$^{-1}$. The 
circle has a radius equal to the virial radius
for this cluster ($10.88'$).}
\label{appz348}
\end{figure*}

\begin{figure*}
\centering
\includegraphics[width=9cm]{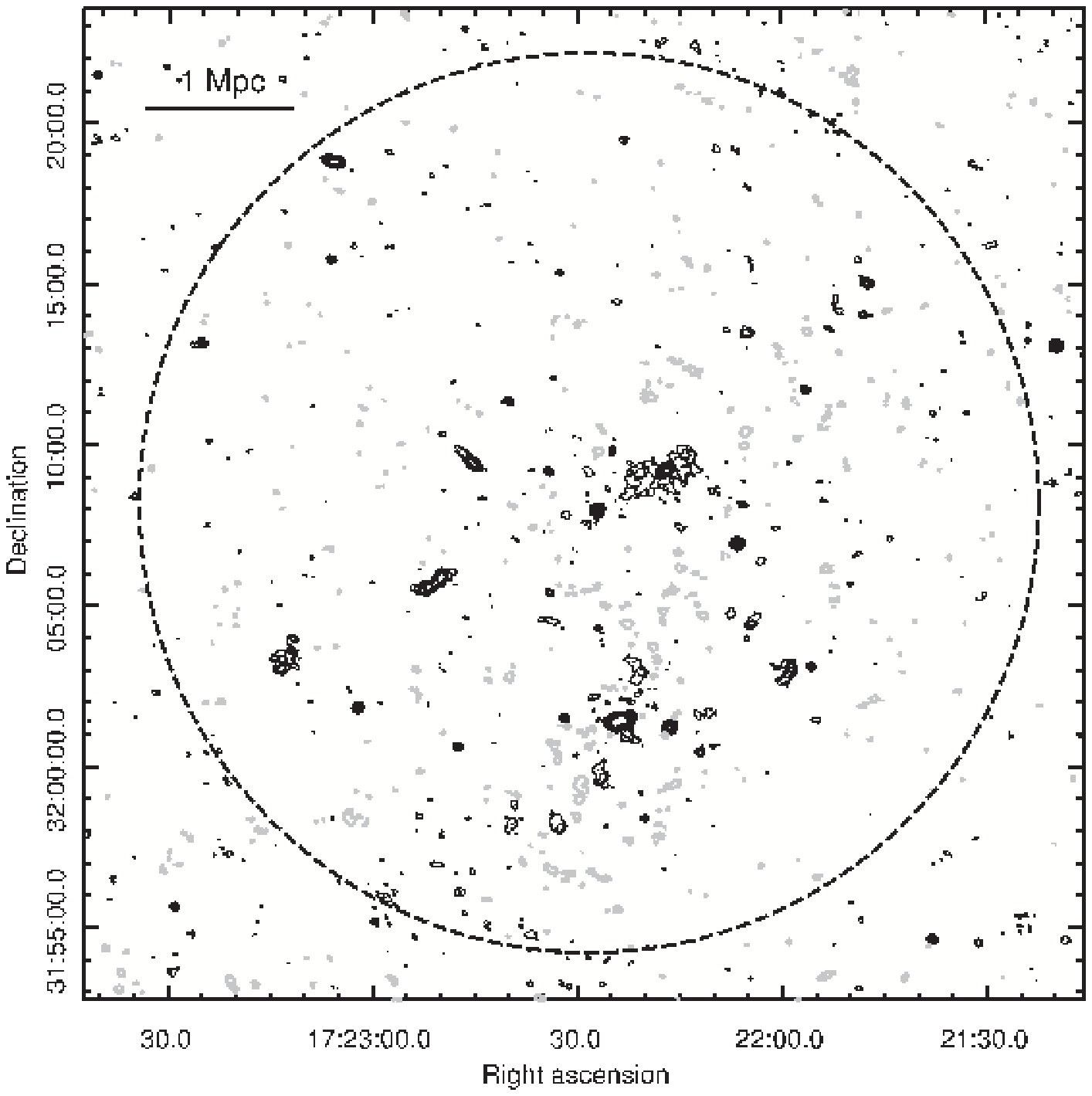}
\includegraphics[width=9cm]{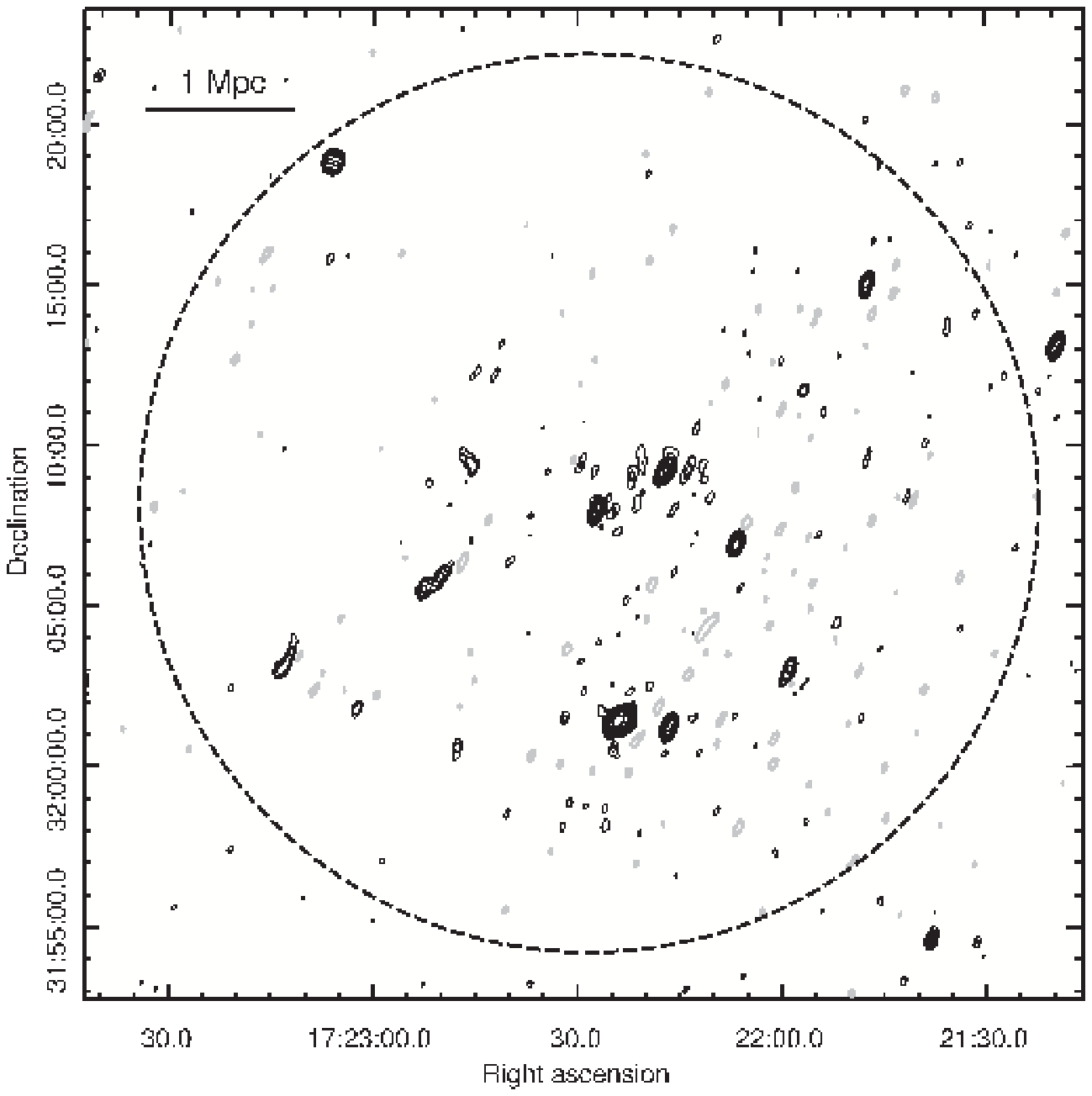}
\caption{{\bf A2261}: GMRT 610 MHz (left) and 235 MHz (right) images 
in 
contours. The
contours are at $0.24\times(\pm1, 2, 4,...)$ mJy beam$^{-1}$ at 610 MHz and 
$1.5\times(\pm1, 2, 4,...)$ mJy beam$^{-1}$ at 235 MHz. The
HPBWs at 610 and 235 MHz are $11.2''\times8.9''$, p. a. $76.0^{\circ}$ and 
$24.4''\times11.8''$, p. a. $-19.0^{\circ}$, respectively. The 
$1\sigma$ levels at 610 and 235 MHz are 0.08 and 0.45 mJy beam$^{-1}$, 
respectively. The circle has a
radius equal to the virial radius
for this cluster ($13.98'$).}
\label{appa2261}
\end{figure*}

\begin{figure*}
\centering
\includegraphics[width=10cm]{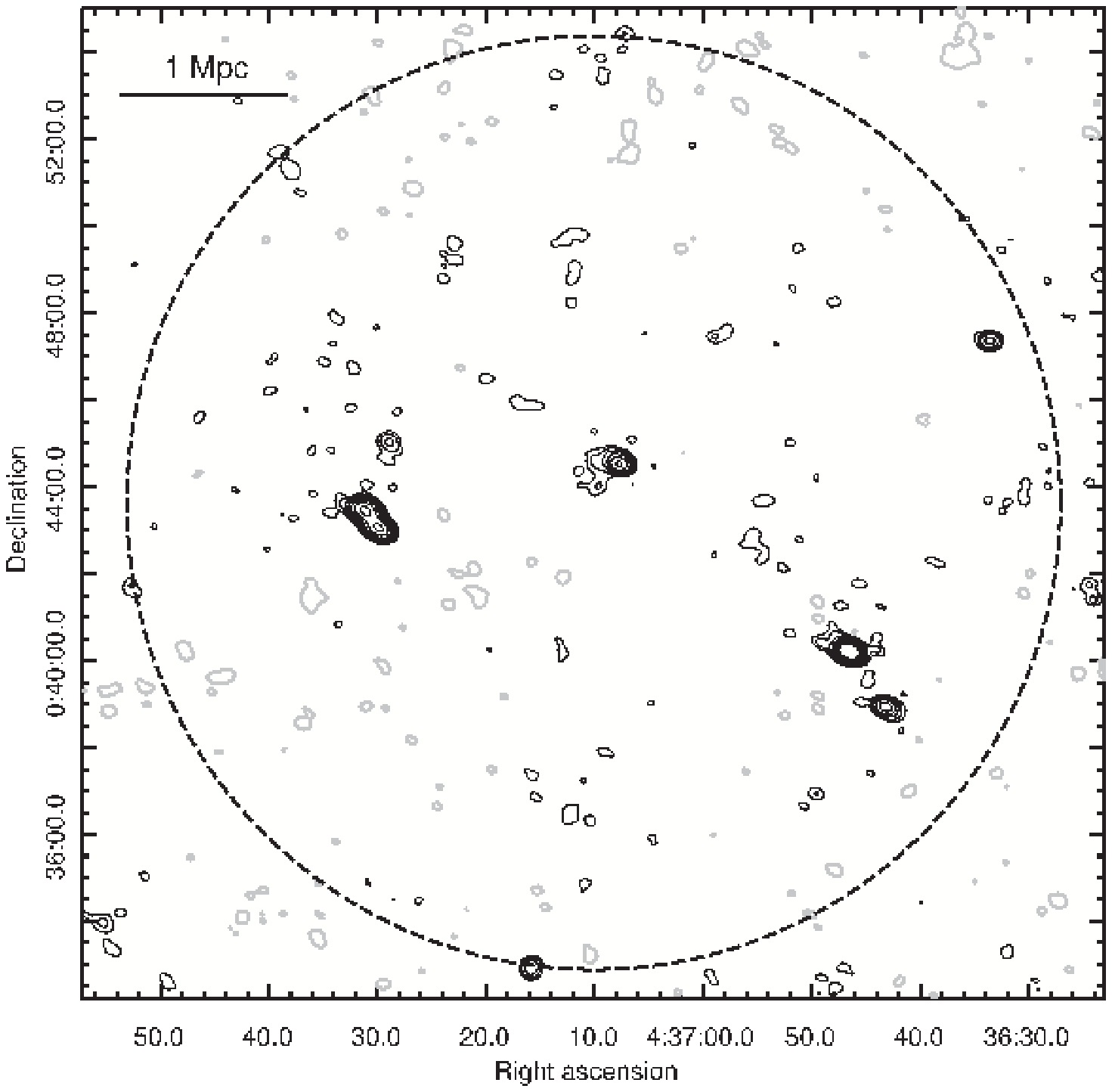}
\caption{{\bf RXCJ0437.1+0043}: GMRT 325 MHz image in contours. The
contours are at $1.0\times(\pm1, 2, 4,...)$ mJy beam$^{-1}$. The HPBW
is $18.6''\times14.3''$, p. a. $82.0^{\circ}$. The 
$1\sigma$ level 
in the image is 0.25 mJy beam$^{-1}$. The circle has a
radius equal to the virial radius
for this cluster ($10.75'$).}
\label{apprx0437}
\end{figure*}

\begin{figure*}
\centering
\includegraphics[width=10cm]{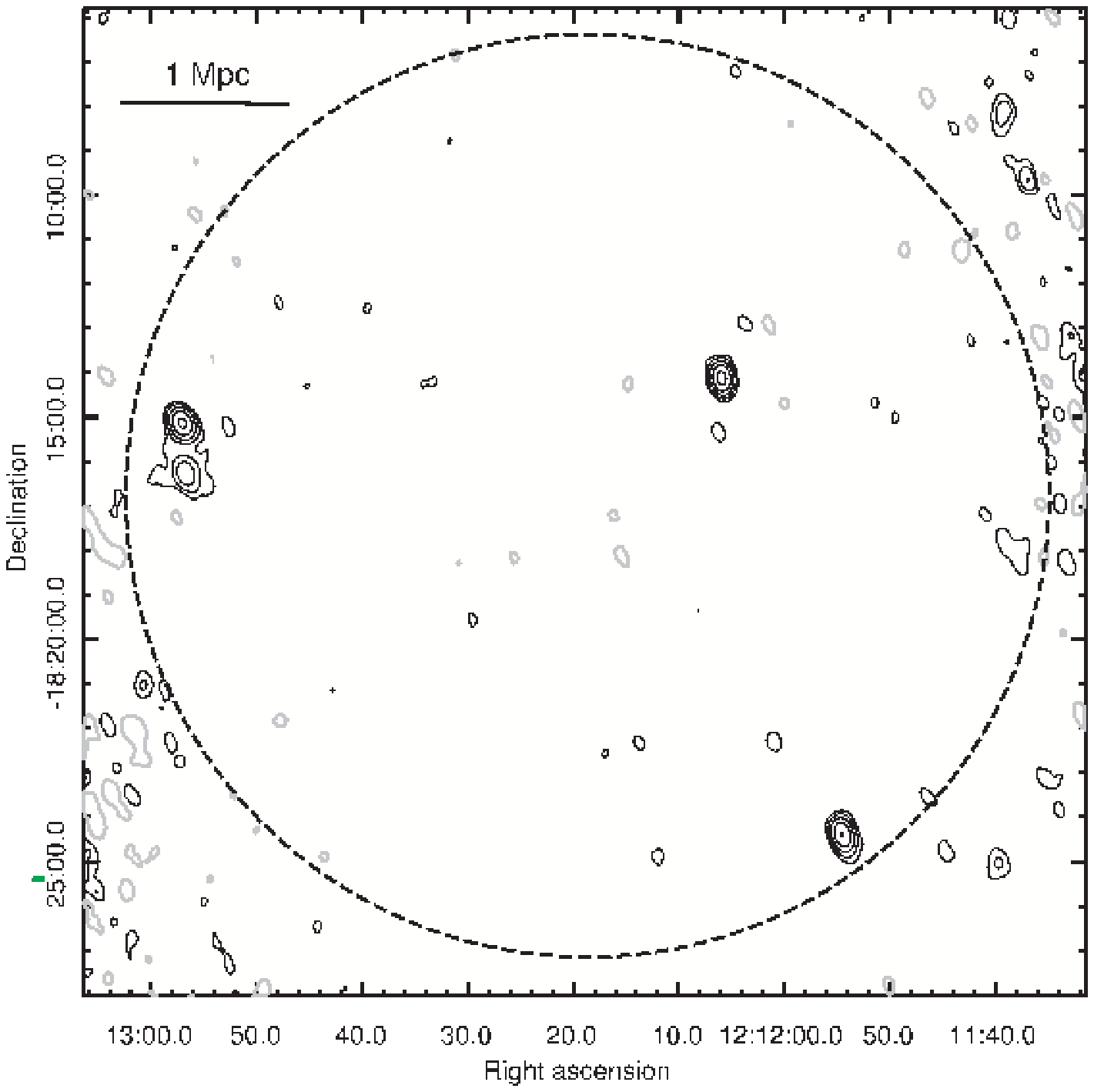}
\caption{{\bf RXCJ1212.3-1816}: GMRT 325 MHz image in contours. The
contours are at $2.0\times(\pm1, 2, 4,...)$ mJy beam$^{-1}$. The HPBW
is $29.6''\times19.7''$, p. a. $10.4^{\circ}$. The 
$1\sigma$ level 
in the image is 0.40 mJy beam$^{-1}$.
The circle has a
radius equal to the virial radius
for this cluster ($10.38'$).}
\label{apprx1212}
\end{figure*}

\begin{figure*}
\centering
\includegraphics[width=10cm]{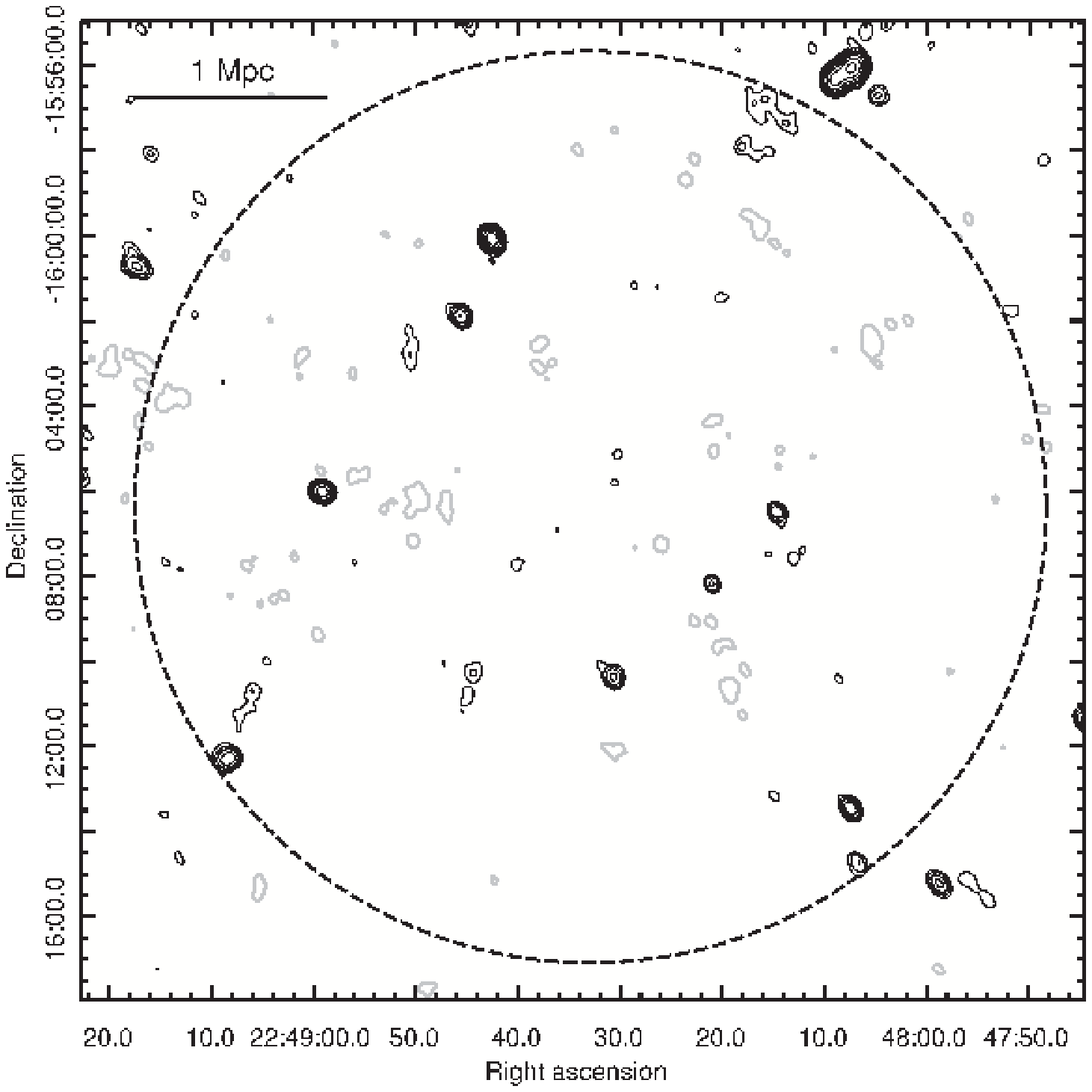}
\caption{{\bf A2485}: GMRT 325 MHz image in contours. The
contours are at $1.0\times(\pm1, 2, 4,...)$ mJy beam$^{-1}$. The HPBW
is $18.2''\times15.4''$, p. a. $36.2^{\circ}$. The 
$1\sigma$ level 
in the image is 0.25 mJy beam$^{-1}$. 
The circle has a
radius equal to the virial radius
for this cluster ($10.71'$).}
\label{appa2485}
\end{figure*}

\end{appendix}

\end{document}